\documentclass[a4paper,fleqn,usenatbib]{mnras}

\usepackage{newtxtext,newtxmath}

\usepackage[T1]{fontenc}
\usepackage{ae,aecompl}

\usepackage{bm}
\usepackage[linesnumbered]{algorithm2e}
\usepackage{color,hyperref}
\usepackage{float}
\usepackage{amsmath}	
\usepackage{amssymb}	
\usepackage{graphicx}	
\usepackage{subcaption}
\usepackage{pdflscape}
\usepackage{afterpage}
\usepackage{colortbl}   
\usepackage[first=0,last=9]{lcg} 

\definecolor{linkcolor}{rgb}{0,0,0.25}
\definecolor{Gray}{gray}{0.9}
\hypersetup{
  colorlinks=true,        
  linkcolor=linkcolor,    
  citecolor=linkcolor,    
  filecolor=linkcolor,    
  urlcolor=linkcolor      
}
\newcommand{\Teff}{T_{\mathrm{eff}}}
\newcommand{\Npts}{$N_{\mathrm{pts}}$}

\title[Chemically tagging birth clusters]{Strong chemical tagging with APOGEE: 21 candidate star clusters that have dissolved across the Milky Way disc}

\author[Price-Jones et. al.]{Natalie Price-Jones$^{1,2}$\thanks{E-mail: price-jones@astro.utoronto.ca}, 
Jo Bovy$^{1,2}$, Jeremy J. Webb$^{1}$, Carlos Allende Prieto$^{3,4}$, \newauthor Rachael Beaton$^{5}$, Joel R. Brownstein$^{6}$, Roger E. Cohen$^{7}$, Katia Cunha$^{8,9}$, \newauthor John Donor$^{10}$, Peter M. Frinchaboy$^{10}$, D. A. Garc\'ia-Hern\'andez$^{3,4}$, Richard R. Lane$^{11, 12}$ \newauthor Steven R. Majewski$^{13}$, David L. Nidever$^{14}$, Alexandre Roman-Lopes$^{15}$
\\
$^{1}$ David A. Dunlap Department of Astronomy and Astrophysics, University of Toronto, 50 St George Street, Toronto ON, M5S 3H4, Canada\\
$^{2}$ Dunlap Institute for Astronomy and Astrophysics, University of Toronto, 50 St. George Street, Toronto, ON M5S 3H4, Canada\\ 
$^{3}$ Instituto de Astrof\'isica de Canarias (IAC), E-38205 La Laguna, Tenerife, Spain\\
$^{4}$ Universidad de La Laguna (ULL), Departamento de Astrof\'isica, E-38206 La Laguna, Tenerife, Spain\\
$^{5}$ The Carnegie Observatories, 813 Santa Barbara Street, Pasadena, CA 91101, USA\\
$^{6}$ Department of Physics and Astronomy, University of Utah, 115 S. 1400 E., Salt Lake City, UT 84112, USA\\
$^{7}$ Space Telescope Science Institute, 3700 San Martin Drive, Baltimore, MD 21218, USA\\
$^{8}$ Steward Observatory, University of Arizona  Tucson AZ 85719\\
$^{9}$ Observat\'orio Nacional/MCTIC, R. Gen. Jos\'e  Cristino, 77,  20921-400, Rio de Janeiro, Brazil\\
$^{10}$ Department of Physics \& Astronomy, Texas Christian University, TCU Box 298840, Fort Worth, TX 76129, USA\\
$^{11}$ Instituto de Astrof\'isica, Pontificia Universidad Cat\'olica de Chile, Av. Vicuna Mackenna 4860, 782-0436 Macul, Santiago, Chile\\
$^{12}$ Instituto de Astronom\'ia y Ciencias Planetarias, Universidad de Atacama, Copayapu 485, Copiap\'o, Chile\\
$^{13}$ Dept. of Astronomy, University of Virginia, Charlottesville, VA 22904-4325, USA\\
$^{14}$ Department of Physics, Montana State University, P.O. Box
173840, Bozeman, MT 59717-3840\\
$^{15}$ Department of Physics \& Astronomy - Universidad de La Serena - Av. Juan Cisternas, 1200 North, La Serena, Chile
}

\date{Accepted XXX. Received YYY; in original form ZZZ}

\pubyear{2020}

\begin{document}
\label{firstpage}
\pagerange{\pageref{firstpage}--\pageref{lastpage}}
\maketitle

\begin{abstract}
	Chemically tagging groups of stars born in the same birth cluster is a major goal of spectroscopic surveys. To investigate the feasibility of such strong chemical tagging, we perform a blind chemical tagging experiment on abundances measured from APOGEE survey spectra. We apply a density-based clustering algorithm to the eight dimensional chemical space defined by [Mg/Fe], [Al/Fe], [Si/Fe], [K/Fe], [Ti/Fe], [Mn/Fe], [Fe/H], and [Ni/Fe], abundances ratios which together span multiple nucleosynthetic channels. In a high quality sample of 182,538 giant stars, we detect twenty-one candidate clusters with more than fifteen members. Our candidate clusters are more chemically homogeneous than a population of non-member stars with similar [Mg/Fe] and [Fe/H], even in abundances not used for tagging. Group members are consistent with having the same age and fall along a single stellar-population track in $\log g$ vs. $\Teff$ space. Each group's members are distributed over multiple kpc, and the spread in their radial and azimuthal actions increases with age. We qualitatively reproduce this increase using $N$-body simulations of cluster dissolution in Galactic potentials that include transient winding spiral arms. Observing our candidate birth clusters with high-resolution spectroscopy in other wavebands to investigate their chemical homogeneity in other nucleosynthetic groups will be essential to confirming the efficacy of strong chemical tagging. Our initially spatially-compact but now widely dispersed candidate clusters will provide novel limits on chemical evolution and orbital diffusion in the Galactic disc, and constraints on star formation in loosely-bound groups.
	\end{abstract} 

\begin{keywords}
Galaxy: structure -- methods: data analysis -- stars: abundances -- stars: statistics -- open clusters and associations: general
\end{keywords}

\section{Introduction}
\label{sec:intro}

The formation and evolution of our Galaxy is challenging to constrain, because we can only observe the present-day snapshot of the Milky Way's current behaviour. Even understanding this current behaviour and the structure of our Galaxy presents significant difficulties, because we must observe it from within its disc. However, increasingly large surveys offer much improved leverage on the problem of understanding the stellar components of the Galaxy as an evolving galactic system. 

One particularly interesting avenue of exploration is that offered by chemical tagging, the process of grouping stars together based on similarity in their chemical abundances \citep{Freeman2002}. These abundances are being measured for hundreds of thousands of stars in current surveys like the RAdial Velocity Experiment (RAVE - \citealt{Steinmetz2006}), the Gaia-ESO survey (GES - \citealt{Gilmore2012}), the Apache Point Galactic Evolution Experiment (APOGEE - \citealt{Majewski2017}), and GALactic Archaeology with Hermes (GALAH - \citealt{DeSilva2015}). Upcoming surveys, such as  those of the 4-metre Multi-Object Spectrograph Telescope (4MOST - \citealt{deJong2016}), the WHT Enhanced Area Velocity Explorer (WEAVE - \citealt{Dalton2012}), the Sloan Digital Sky Survey V (SDSS-V -  \citealt{Kollmeier2017}), and the MaunaKea Spectroscopic Explorer (MSE -\citealt{MSE2019}), will further extend these abundance measurements. Increasing the number of stars with precise chemical measurements is a crucial requirement for chemical tagging on all scales \citep{Ting2015a}.

Weak chemical tagging has been used to distinguish populations of stars belonging to different components of the Galaxy, like the disc, halo, and bar (e.g. \citealt{Bovy2012, Anders2015, Hawkins2015, Wojno2016}). However, one of the most tantalizing prospects of chemical tagging is its application in the strong limit: using it to identify `birth clusters'. Birth clusters are groups of stars that formed from the same giant molecular cloud (GMC) but have since dispersed in phase space such that they are no longer recognizable as a stellar association like a standard open cluster. Most groups of stars forming in the same cloud are expected to disperse in less than 100 Myr \citep{Lada2003}, but open clusters are a more tightly bound exception. Because of this, they are often taken as a proxy for birth clusters, as they can still be identified as overdensities in phase space. However, finding the stellar siblings that originated in the same birth cluster but have since dispersed requires a chemical tagging approach. Chemical tagging relies on very high precision and accuracy in measurements of stellar chemistry, as well as two crucial assumptions. The first of these is that stars from the same birth cluster are chemically homogeneous, while the second is that each birth cluster has a sufficiently unique chemical signature that they do not overlap in chemical space.

Investigating the validity of these assumptions has been the focus of many recent works. The assumptions of birth cluster homogeneity and uniqueness are most often tested on open clusters in their role as birth cluster proxies. Observational tests have established that open clusters are chemically homogeneous below the level of measurement uncertainties in large surveys (e.g. \citealt{DeSilva2006, DeSilva2007, Bovy2016}), and these results are further reinforced by checks on wide binaries, expected to be homogeneous for the same reason \citep{Hawkins2020}. Some focused studies of open cluster M67, (e.g. \citealt{Souto2018,Souto2019,Liu2019}) have identified chemical differences between its members, but these apparent discrepancies can be attributed to the different ways atomic diffusion influences surface abundances at various stages of stellar evolution \citep{Dotter2017}, and evolved stars in M67 were still found to be chemically homogeneous with each other.

While the homogeneity of open clusters has been well established, determining whether they possess unique chemical signatures has proven more challenging. In \citet{Ting2015a}, the authors found that the chemical space represented by abundances measured in current surveys would be difficult to fully sample. However, \citet{Price-Jones2018} demonstrated that using stellar spectra extends the dimensionality of chemical space. Tests attempting to distinguish the chemical signatures of open clusters in abundance space have noted difficulty in separating member of different clusters (e.g. \citealt{Blanco-Cuaresma2015}), perhaps limiting chemical tagging to the identification of families of clusters with similar ages \citep{Garcia-Dias2019}. In addition, the work of \citet{Ness2018} identifying APOGEE field stars with the same chemical signature as open cluster members further highlights the difficulty of distinguishing birth cluster signatures. Follow up work in a similar vein in \citet{Ness2019} would seem to indicate that detailed chemistry is a deterministic property of age and [Fe/H] that does not necessarily change with birth location. However, in \citet{Price-Jones2019}, we show that simulated birth clusters can be chemically tagged even when their chemical signatures are given by randomly selected APOGEE stars, essentially placing no requirement for chemical signature uniqueness.

Despite the questions still surrounding the feasibility of using strong chemical tagging to find birth clusters, the technique has already been employed on chemical spaces from surveys of all sizes. There have been numerous attempts at blind chemical tagging, where only chemical information is used to determine group membership. The first example of blind chemical tagging in \citet{Mitschang2014} highlighted the challenges of the technique, as that work identified groups that were born at the same time (co-eval), but not necessarily also in the same place (co-natal). In \citet{Blanco-Cuaresma2015}, the blind attempt to reassign open cluster members to the correct cluster resulted in final clusters that were aggregates of several open clusters. However, subsequent attempts, using both real and simulated chemical spaces, have been promising. \citet{Hogg2016} was able to recover known clusters as well as new stellar associations using APOGEE data, and in our APOGEE-like simulated chemical space described in \citet{Price-Jones2019} we successfully chemically tagged more than 30\% of our input clusters.

In this work, we blindly chemically tag the APOGEE chemical space using a density-based clustering algorithm. Unlike some previous attempts, our approach requires no assumption about the number of groups we expect to find. The algorithm we choose, Density-Based Spatial Clustering Applications with Noise (DBSCAN), identifies groups as overdensities in chemical space, but does not assign every star to a group, as it is empowered to flag some stars as `noise'. This flagging of noise stars is especially helpful given our expectation that many APOGEE stars will be the sole representative of their birth cluster, given that the sample that is a relatively small fraction of all stars in the Galaxy. We analyze the groups of chemically homogeneous stars identified by DBSCAN to confirm their homogeneity in all abundances measured by APOGEE. In addition, we investigate stellar ages and kinematics to assess whether our groups are consistent with being stellar birth clusters.

This work is organized as follows. In \S\ref{sec:data}, we describe the APOGEE survey and the chemical space we use for the chemical tagging, as well as kinematic and age measurements for our stars. We follow this in \S\ref{sec:chemicaltagging} with a description of the DBSCAN algorithm, as well as an explanation of how we choose our chemical space and how we apply DBSCAN to APOGEE. \S\ref{sec:clusters} outlines the properties of the groups identified by DBSCAN, both in chemical space and in terms of stellar ages. In \S\ref{sec:dyn}, we consider the orbits of the stars identified as group members, and compare the properties of these orbits to $N$-body cluster simulations of cluster dissolution a realistic potential. We discuss the plausibility of interpreting our groups as stellar birth clusters in \S\ref{sec:discussion}, and summarize our conclusions in \S\ref{sec:conclusions}.

\section{Data}
\label{sec:data}

To perform chemical tagging, a large amount of chemical data is required. We use elemental abundances from the APOGEE spectroscopic survey to investigate clustering in chemical space. We supplement this with kinematic information from the second data release from the \emph{Gaia} mission (\citealt{GaiaCollaboration2016}, \citealt{GaiaCollaboration2018}) to further investigate the possibility that the stars identified as belonging to the same group could have formed in the same location.

\subsection{Stellar Chemistry from APOGEE}
\label{sec:chemistry}
This analysis makes use of data from APOGEE \citep{Majewski2017}, the SDSS-IV \citep{Blanton2017} survey focused on measuring stellar chemistry for a large sample of stars across  the Milky Way. APOGEE uses two high resolution ($R\sim 22,500$) $H$-band spectrographs, with one in each hemisphere to obtain full-sky coverage. The original APOGEE spectrograph is mounted on the 2.5-m Sloan Foundation telescope at Apache Point Observatory \citep{Gunn2006}. Another spectrograph was added in 2017 to observe the Southern Sky from the 2.5-m Ir\'en\'ee du Pont Telescope at Las Campanas Observatory \citep{Bowen1973}. Both instruments are fibre-fed, observing 300 targets simultaneously \citep{Wilson2019}. APOGEE's sixteenth data release (DR16; \citealt{Ahumada2019}, J\"onsson et al., in prep) is the first data release to consist of observations from both telescopes. Except for stars in a number of fields towards the bulge and Galactic centre, each of the APOGEE survey's primarily red giant targets \citep{Zasowski2013,Zasowski2017} is observed in at least three separate visits. The individual visit spectra are processed through a pipeline that performs radial velocity correction before combining the visits into a single spectrum \citep{Nidever2015}. The combined spectra are passed through the APOGEE Stellar Parameter and Chemical Abundances Pipeline (ASPCAP; \citealt{GarciaPerez2016}), which measures the effective temperature ($\Teff$), surface gravity ($\log g$), and the abundances of 24 elements (C, N, O, Na, Mg, Al, Si, P, S, K, Ca, Ti, V, Cr, Mn, Fe, Co, Ni, Cu, Ge, Rb, Ce, Nd, and Yb). 

In this work, we make use of elemental abundances determined by \texttt{astroNN} \citep{Leung2019}, a neural network that was trained on the results of ASPCAP for APOGEE's DR14 \citep{Abolfathi2018, Holtzman2018} for high signal to noise ratio (SNR) spectra. The network produces higher precision abundances with reliable results even for stars with a lower SNR than APOGEE's nominal target of SNR=100. The \texttt{astroNN} produces the following abundances, as it is these abundances for which ASPCAP is most reliable: [C/Fe], [N/Fe], [O/Fe], [Na/Fe], [Mg/Fe], [Al/Fe], [Si/Fe], [S/Fe], [K/Fe], [Ca/Fe], [Ti/Fe] (derived from Ti I lines), [V/Fe], [Mn/Fe], [Fe/H], [Ni/Fe]. These results are published in an SDSS Value Added Catalogue (VAC)\footnote{\url{https://data.sdss.org/sas/dr16/apogee/vac/apogee-astronn/}}. Though this VAC initially contains 473,307 stars, we disregard duplicate entries, stars without de-reddened magnitude measurements, and stars observed during the commissioning of the instrument. This reduces the number of stars considered to 414,631.

In Table~\ref{tab:props} we list the abundances that were measured for each star, as well as other stellar properties on which we make quality cuts. We start by only considering stars with overall signal to noise ratios greater than 50. We restrict ourselves to stars with temperatures between 3500 K and 5000 K, as this is the range over which ASPCAP results (and consequently \texttt{astroNN}) are most reliable, and we require that temperature measurements have less than 100 K uncertainty. We further limit our consideration to stars with $0 < \log g < 4$, and require that their surface gravity uncertainty be less than 0.2 dex. Stars with higher surface gravity uncertainties tend to be dwarf stars, and those high uncertainties mean their abundance measurements are of poorer quality. Our combined limits on $\log g$, $T_{\rm eff}$, and their uncertainties ensure we are only considering giant or subgiant stars for our analysis. These restrictions reduce our original sample of 414,631 stars to 201,755. Within this reduced sample, we also make cuts related to the abundances we use for chemical tagging, requiring their uncertainties be less than 0.15 dex. The abundances chosen for chemical tagging and their associated quality cuts are explained in more detail in \S\ref{sec:sim}.

\subsection{Stellar Kinematics and Ages}
\label{sec:kinematics}
In order to further analyze the groups of stars we chemically tag, we employ kinematic measurements to determine whether their orbits are consistent with having been born in the same cluster. For each star in our APOGEE sample, limited by the constraints described in the previous section, we collect right ascension, declination, and their respective proper motions from the \emph{Gaia} data release 2 (DR2) catalogue \citep{GaiaCollaboration2018}. We take line-of-sight velocities from APOGEE, because they are higher precision than those obtained by \emph{Gaia}, even where \emph{Gaia} has measured the line-of-sight velocity. We use distances from the \texttt{astroNN} VAC, which are a weighted combination of the distance estimated from \emph{Gaia} parallax and that inferred by machine learning from APOGEE spectra and 2MASS photometry \citep{Leung2019a}. Since these properties will not be used for chemical tagging, we do not restrict our sample based on their quality.

With the kinematic information listed above, we compute orbital actions for each star, where we assume the Milky Way potential is the \texttt{MWPotential2014} described in \citet{Bovy2015a}. Furthermore, we assume the Sun is at a radius of 8 kpc \citep{Bland-Hawthorn2016}, 20.8 pc above the plane \citep{Bennett2019}, with $v_x$, $v_y$, and $v_z$ equal to $-11.1$, $232.24$, and $7.25$ km s$^{-1}$ respectively (\citealt{Schonrich2010, Bovy2012a}). These parameters are used throughout our work for any transformation between heliocentric and Galactocentric coordinates. Actions are calculated with the \texttt{galpy} Python package \citep{Bovy2015a} using the St\"ackel approximation \citep{Binney2012}. These actions give an overall summary of the orbit of any given star in our sample, reducing our 6D phase space down to three dimensions: $J_{\rm R}$, $J_{\phi}$, and $J_{\rm z}$.

We also use age estimates for each of the stars in our sample from \citet{Mackereth2019} (these ages are included in the \texttt{astroNN} VAC). The stellar ages are determined by training a Bayesian convolutional neural network on a set of APOGEE spectra and their corresponding ages (determined with high precision using asteroseismology). There is an expected relationship between a star's age and its APOGEE spectrum due to the presence of carbon and nitrogen lines in the waveband, as these elements can be related to stellar mass (and therefore age, e.g. \citealt{Masseron2015}). Having learned the relationship between spectrum and age for the training set, the neural network estimated ages for the remaining APOGEE stars from their spectra. These ages have typical uncertainties of 30-35\%, or about 1.5 Gyr for a star predicted to have an age of 5 Gyr.

\begin{table}
\centering
\caption{Measured stellar properties for stars in our sample, with any related constraints. Rows highlighted in grey correspond to elements used to compose the chemical space that we use for the clustering analysis (see \S\ref{sec:sim}). Our final chemical space is eight dimensional, with our initial sample of $414,631$ stars reduced to just $182,538$ after combining all constraints listed below.}
	\begin{tabular}{r|l}
	\label{tab:props}
		property &  constraint  \\
		\hline
		\textbf{ASPCAP} & \\
		{Spectrum SNR}       & $\mathrm{SNR} > 50$ \\ 
		\hline
		\texttt{astroNN} & \\
		{$\Teff$}   & $3500 \mathrm{K} < \Teff < 5000 \mathrm{K}$ \\
		            & and $\sigma_{\Teff}<$ 100 K  \\
		{$\log g$}  & $0 < \log g < 4$ \\
					& and $\sigma_{\log g} <$ 0.2  \\
		{[C/Fe]}    & None   \\
		{[N/Fe]}    & None \\
		{[O/Fe]}    & None \\
		{[Na/Fe]}   & None  \\
		\rowcolor{Gray}
		{[Mg/Fe]}  & $\sigma_{\mathrm{[Mg/Fe]}}< 0.15$ dex  \\
		\rowcolor{Gray}
		{[Al/Fe]}  & $\sigma_{\mathrm{[Al/Fe]}}< 0.15$ dex  \\
		\rowcolor{Gray}
		{[Si/Fe]}   & $\sigma_{\mathrm{[Si/Fe]}}< 0.15$ dex \\
		{[S/Fe]}    & None  \\
		\rowcolor{Gray}
		{[K/Fe]}    &  $\sigma_{\mathrm{[K/Fe]}}< 0.15$ dex \\
		{[Ca/Fe]}   & None  \\
		\rowcolor{Gray}
		{[Ti/Fe]}   & $\sigma_{\mathrm{[Ti/Fe]}}< 0.15$ dex  \\
		{[V/Fe]}    & None \\
		\rowcolor{Gray}
		{[Mn/Fe]}   & $\sigma_{\mathrm{[Mn/Fe]}}< 0.15$ dex\\
		\rowcolor{Gray}
		{[Fe/H]}    & $\sigma_{\mathrm{[Fe/H]}}< 0.15$ dex \\
		\rowcolor{Gray}
		{[Ni/Fe]}   & $\sigma_{\mathrm{[Ni/Fe]}}< 0.15$ dex \\

	\end{tabular}	
\end{table}

\section{Chemical tagging the APOGEE data with DBSCAN}
\label{sec:chemicaltagging}

\subsection{DBSCAN}
\label{sec:dbscan}

Density-based spatial applications with noise (DBSCAN,  \citealt{Ester1996}) uses the density of measurements to find groups in data. In our subsequent discussion we will always refer to the associations found by DBSCAN as `groups' or `cluster candidates' rather than the more common `clusters'. We will reserve `clusters' to refer to stellar associations whose members are known to have been born in the same place; the groups found by DBSCAN are merely candidates for this cluster designation.

The DBSCAN algorithm identifies stars as belonging to groups by  first assessing the chemical space around each star and deciding whether there are enough neighbouring stars to consider the region high-density. DBSCAN's two parameters, $\epsilon$ and \Npts, determine the size of a star's chemical neighbourhood and the required number of stars for a high-density designation. $\epsilon$ is the radius of the chemical space hypersphere that is considered the chemical neighbourhood of each star. If there are at least \Npts\,  stars inside a hypersphere with radius $\epsilon$  that is centred on a particular star, then that star will be designated a `core star'. For convenience, we will refer to the hypersphere with radius $\epsilon$ centred on a star as that star's `chemical neighbourhood'. 

A star that is in the chemical neighbourhood of a core star but does not have at least \Npts\, stars in its own chemical neighbourhood is designated a `border star'. Any star that does not receive either of the above designations is considered a `noise star', and will not be assigned to a group by the algorithm. The noise star designation is very useful when considering stars that are the only representative of their birth cluster in our sample, allowing us to discard them and avoid contaminating groups that may represent true birth clusters.
 The DBSCAN algorithm proceeds by first checking every star to see if it satisfies the criteria to be a core star. Once all core stars have been identified, DBSCAN designates the remaining stars as either `border' or `noise' stars. Core stars are then  re-examined, and those with overlapping chemical neighbourhoods are merged into larger groups, along with their attendant border stars.  Once all stars associated with overlapping core stars have been merged into groups, DBSCAN returns an integer label for each star to indicate which group it belongs to, giving a special flag for noise stars. For an extended description of this algorithm, see \citet{Price-Jones2019}.

This density based approach means we are not required to know a priori how many groups we expect to find in our data, making it very useful for blind application to a large chemical space. 

\subsection{Choosing a chemical space with simulations}
\label{sec:sim}

\begin{figure}
\centering
\includegraphics[width = \linewidth]{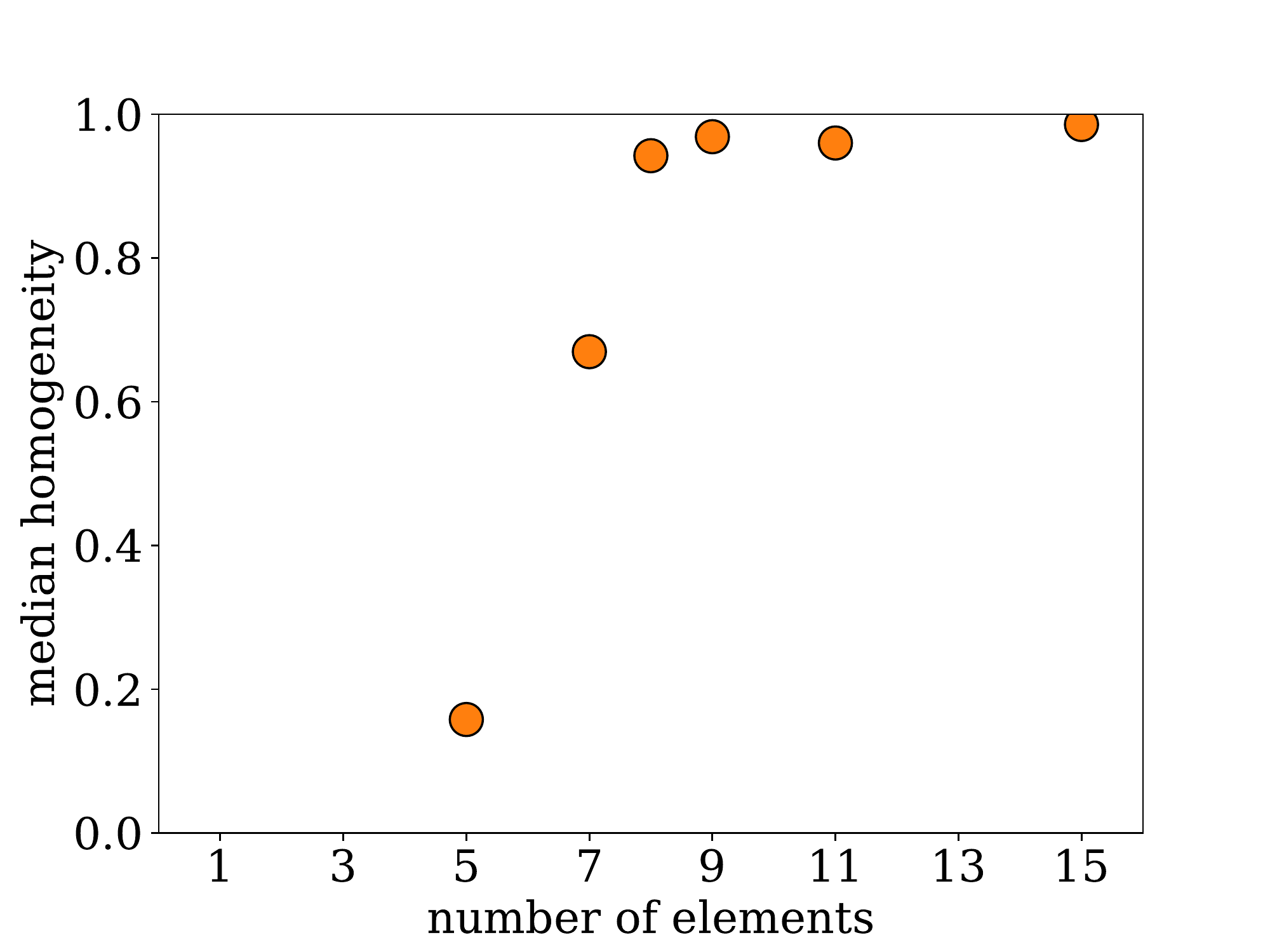}
\caption{Median group homogeneity as a function of the number of elements used for chemical tagging in our simulated chemical space. The initial group of five elements consists of [Mg/Fe], [Al/Fe],[Si/Fe], [Mn/Fe], and [Fe/H]. As the number of elements is increased, we add [Ti/Fe] and [Ni/Fe], then [K/Fe], then [C/Fe] and [N/Fe], and finally [Na/Fe], [S/Fe], [Ca/Fe] and [V/Fe] to build up from five to fifteen elements.}
\label{fig:homog}
\end{figure}

To select a chemical space for our APOGEE sample, we first estimate the ability of DBSCAN to recover groups in different chemical spaces. We investigated this question in detail in \citet{Price-Jones2019}, and follow the procedure of that work to create a simulated chemical space to closely mimic our APOGEE sample. We summarize this process below.

We start by taking the simulated survey volume to be an annulus with a width of 6 kpc, with its inner edge 5 kpc from the Galactic centre. We further assume that the stellar mass density of the Milky Way is 0.05 $M_{\odot}\,\mathrm{pc}^{-3}$ everywhere \citep{Bovy2017}. We use our survey volume and the mass density to find the total stellar mass accessible by our simulated survey. Assuming a Kroupa initial mass function \citep{Kroupa2001}, we find that stars in our survey volume have an average mass of 0.6 $M_{\odot}$, which allows us to make a simple approximation of the total number of stars in the region (approximately 26 billion).

We set the number of stars `observed' by our simulated survey to be equal to the number of stars in our APOGEE sample. This number depends on the chemical space we choose to test on, because we require for our APOGEE sample that stars have uncertainty less than 0.15 dex in all abundances used for tagging. Thus for each of the chemical spaces described, we determine the number of stars in our simulated space by first finding the number of APOGEE stars that meet this uncertainty requirement. We begin by working with a five-dimensional chemical space with axes that correspond to [Mg/Fe], [Al/Fe], [Si/Fe], [Mn/Fe], and [Fe/H], which after our quality cuts on the abundance uncertainties, is populated by 186,879 stars. We then test a seven-dimensional chemical space with the aforementioned elements, as well as [Ti/Fe] and [Ni/Fe] (186,762 stars). We add [K/Fe] to create an eight-dimensional chemical space with 182,538 stars. To test a nine-dimensional chemical space, we add [O/Fe] (182,472 stars), and follow this by testing an eleven-dimensional space by adding [C/Fe] and [N/Fe] (182,037 stars). Finally, we test on the full slate of fifteen abundances, including [Na/Fe], [S/Fe], [Ca/Fe], and [V/Fe] with all abundances previously listed, which reduces the total number of stars to 144,199. Though we could have made alternate choices for the abundances to compose each abundance space with a lower dimensionality than fifteen, as the dimensionality decreases, the number of possible combinations of elements swiftly becomes too large to effectively test. In addition, we want our chemical space to contain a variety of nucleosynthetic pathways. We provide a more detailed rationale for the abundance ratios we choose in \S\ref{sec:choosing}. 

For a given chemical space, our simulated survey has an overall sampling fraction of approximately $7\times 10^{-6}$ (or $5\times 10^{-6}$ when using all fifteen abundances for the chemical space). We assume our clusters follow a power law cluster mass function (CMF) with an exponent of $-2.1$, a lower mass limit of $50 M_{\odot}$ and an upper mass limit of $10^7 M_{\odot}$ \citep{Ting2015a}.

We draw a cluster mass from this distribution and apply the sampling fraction to determine the number of stars actually observed by our simulated survey. Rather than a blanket multiplication of the overall sampling fraction for every cluster, we use the overall sampling fraction to characterize an exponential distribution from which we draw a unique sampling fraction for each cluster. This approach to the sampling fraction allows us to have some clusters sampled more rigorously than others, as we expect is true for the real data. We repeat this process of drawing cluster masses and applying unique sampling fractions until we have accumulated enough stars to meet our total required number. As described above, this number varies depending on the number of abundances we use for chemical tagging. 

\begin{figure*}
\centering
\includegraphics[width = \linewidth]{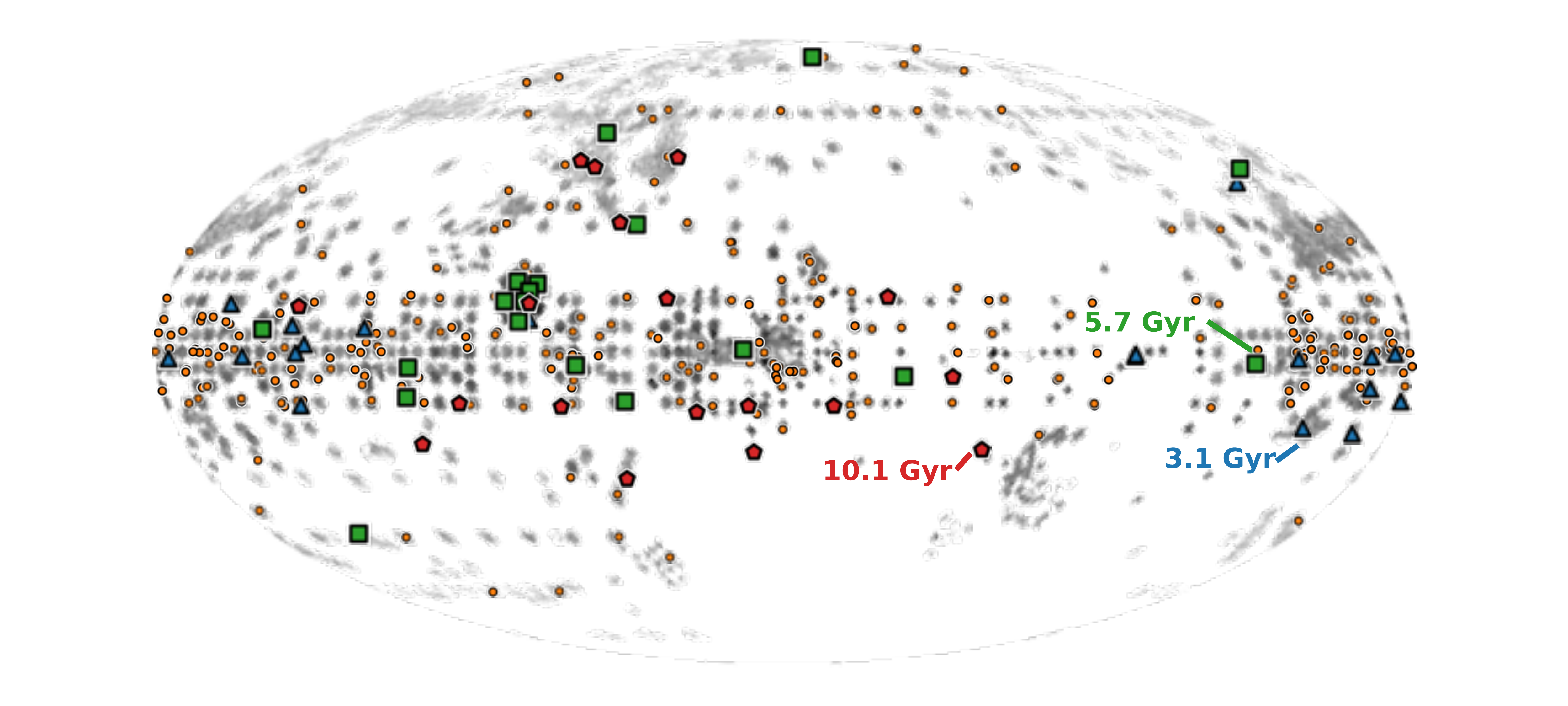}
\caption{ The members of our groups shown in Galactic coordinates (orange circles), with the full sample considered for chemical tagging shown as a histogram in the background. The oldest, median age, and youngest groups are shown with red pentagons, green squares, and blue triangles respectively. As is typical for our groups, the members of these three are spread across the entire APOGEE observational footprint, but our youngest group is the most confined to the Galactic mid-plane, while the oldest has the most stars at high Galactic latitude.}
\label{fig:onsky}
\end{figure*}

Once we have determined how the stars required for each chemical space are divided into clusters that obey a CMF and our overall sampling fraction, we draw for each cluster a random star from a subset of stars in APOGEE's DR16. To ensure realistic chemical signatures, the randomly selected stars must have no \textsc{APOGEE\_STARFLAG} flag set and a measurement for surface gravity as well as all fifteen abundances reported by \texttt{astroNN}. These stars provide the median chemical signature for the cluster. We then generate observed chemical signatures for each star in cluster by adding to the median signature an error in each abundance drawn from a normal distribution characterized by the \texttt{astroNN} uncertainty at 50 SNR for that element (see the `optimistic' column of Table 1 in \citealt{Price-Jones2019}). This emulates the effect of observing these stars by introducing variation of abundances within in the same cluster due to measurement error.

Each of our simulated surveys now consist of the same number of stars we would use when applying DBSCAN to the real data restricted to the same chemical space, and each of those stars has a fifteen element chemical signature. In addition, each star has a true cluster assignment, allowing us to test DBSCAN's performance.

The simulated chemical spaces permit a comparison of the performance of DBSCAN with various choices made in defining the space to chemically tag. In Figure~\ref{fig:homog}, we show the median homogeneity of the recovered groups as a function of the number of elements used for chemical tagging. The homogeneity describes the degree to which the groups are dominated by a single input cluster, which we want to be as high as possible. A median homogeneity value of one in Figure~\ref{fig:homog} indicates that all groups found by DBSCAN consist entirely of members of a single input cluster, while a value near zero indicates that the groups are composed of members from a large array of input clusters. Based on Figure~\ref{fig:homog}, we choose to proceed with an eight-dimensional chemical space, as this choice gives a median homogeneity score higher than 0.9 in our simulations while still leaving multiple abundance ratios out of the tagging process to serve as a check on chemical homogeneity of the found groups.

\subsection{Blind Chemical Tagging with APOGEE}
\label{sec:choosing}

Based on the results of the simulations described in the previous section, we select eight elements to define our chemical space: [Mg/Fe], [Al/Fe], [Si/Fe], [K/Fe], [Ti/Fe], [Mn/Fe], [Fe/H], and [Ni/Fe]. We require that stars have uncertainties less than 0.15 dex for each of the eight elements we choose. This allows us to retain 182,538 stars from our sample of 201,755 giant stars with high SNR and good $\Teff$ and $\log g$ measurements. Our choice of eight dimensions corresponds to the approximately ten dimensions we expect for an APOGEE chemical space \citep{Price-Jones2018}. This choice is also partially motivated by the challenge of ensuring high-quality abundance data for all stars. As the number of dimensions increases, the number of stars with abundance uncertainty < 0.15 dex in each dimension decreases. We select our eight abundance ratios based on their ability to probe different nucleosynthetic pathways. We choose [Mg/Fe] as a quintessential example of $\alpha$-element abundance to probe the contribution of type II supernovae (SN II) enrichment to chemical evolution prior to a star's birth. [Si/Fe] serves as a complement to this, being mostly produced in SN II with a small contribution from SN Ia. [Al/Fe] and [K/Fe] represent the odd-Z elements; while the source of Al is SN II (with some metallicity dependence; \citealt{Nomoto2013}), the evolution of K is less clear. [Ti/Fe] is another abundance that is thought to be primarily enhanced by SN II, but the possible contribution of SN Ia makes it an interesting inclusion. [Mn/Fe], [Fe/H], and [Ni/Fe] represent the iron-peak abundances, produced primarily in SN Ia with additional contribution from SN II. 

Our chosen elements have well-defined lines in the APOGEE waveband, and so are typically well measured. We eschew using C or N, as surface abundances of these elements are known to evolve due to internal mixing processes as stars progress along the giant branch \citep{Masseron2015}. We reject [Na/Fe], [S/Fe], and [V/Fe] on the basis that these elements are challenging for APOGEE to measure (\citealt{Majewski2017}, J\"onsson et al. in prep). The remaining elements, [O/Fe] and [Ca/Fe], we reserve as a way to confirm that identified groups are chemically homogeneous across all abundances, and not just the ones we select for tagging.

Having selected our chemical space, we choose a Euclidean metric to measure distances. We apply DBSCAN according to our results in \citet{Price-Jones2019}, where we found that we recovered the simulated input clusters most reliably when we chose $\epsilon$ and \Npts\, that maximized the total number of groups found by DBSCAN. Maximizing the number of groups identified is desirable when trying to get the most out of a chemical space. Furthermore, in \citet{Price-Jones2019} we showed that making this choice of $\epsilon$ and \Npts\, also results in the recovered groups having high homogeneity, such that each was dominated by members of a single input cluster. In addition to using homogeneity scores to assess how the groups matched input clusters, tests were performed to measure the total fraction of input clusters recovered for the different $\epsilon$ and \Npts\, values. This measurement of recovery fraction was done by randomly choosing ten input clusters of sufficient size to be detected by DBSCAN and checking whether each of those clusters corresponded to a group identified by DBSCAN. This process was repeated several times to generate a robust `recovery fraction' statistic. The values of $\epsilon$ and \Npts\, that maximized this statistic also maximized the number of recovered groups, and each of those groups was strongly dominated by members of a single input cluster. This relationship between recovery fraction and number of groups was evident regardless of the uncertainty on the abundance values chosen for our simulated dataset. 

Motivated by the relationship between the number of groups identified by DBSCAN and the fraction of input clusters recovered, we test several values of $\epsilon$ and \Npts\, when applying DBSCAN to our subset of the APOGEE data. In this case, we find that that $\epsilon=0.02$ dex and \Npts$=3$ result in the most groups identified by DBSCAN. Since the simulated experiments summarized above indicate these $\epsilon$ and \Npts\, values will result in the highest fraction of input clusters recovered, we investigate the stellar groups identified by DBSCAN with these parameter values.

Initially, DBSCAN identifies 2,762 groups in our chemical space. We restrict the subsequent investigation to groups with more than 15 members, providing us with enough stars per group to be reasonably confident in our statistical inferences about group properties. We further demand that good candidate clusters are more compact than their surroundings by using a cut on the \emph{silhouette coefficient}. The silhouette coefficient is a commonly employed metric in clustering studies. It is defined for star $i$ in our sample as follows:

\begin{equation}
	S_i = \frac{d_{\mathrm{intra}}-d_{\mathrm{inter}}}{\max(d_{\mathrm{intra}},d_{\mathrm{inter}})},
\end{equation}
where $d_{\mathrm{intra}}$ is the mean distance between star $i$ and all other members of its group, and $d_{\mathrm{inter}}$ is the mean distance between star $i$ and all members of the nearest group in chemical space. $S_i$ can take values from $-1$ to 1, with 1 indicating the star is closer to members of its group than the neighbouring group, 0 indicating the star overlaps with members of the neighbouring group, and $-1$ indicating that the star is overall closer to the neighbouring group than other members of its own group. To get a value of the overall silhouette coefficient $S^g$ for each group, we compute the mean of the $S_i$'s across all of the the member stars. With an $S^g$ for each group, we restrict our subsequent analysis to the groups with $S^g>0$. This ensures we only consider groups for which members were typically closer to each other than to stars in other groups.

\begin{figure}
\centering
\includegraphics[width = \linewidth]{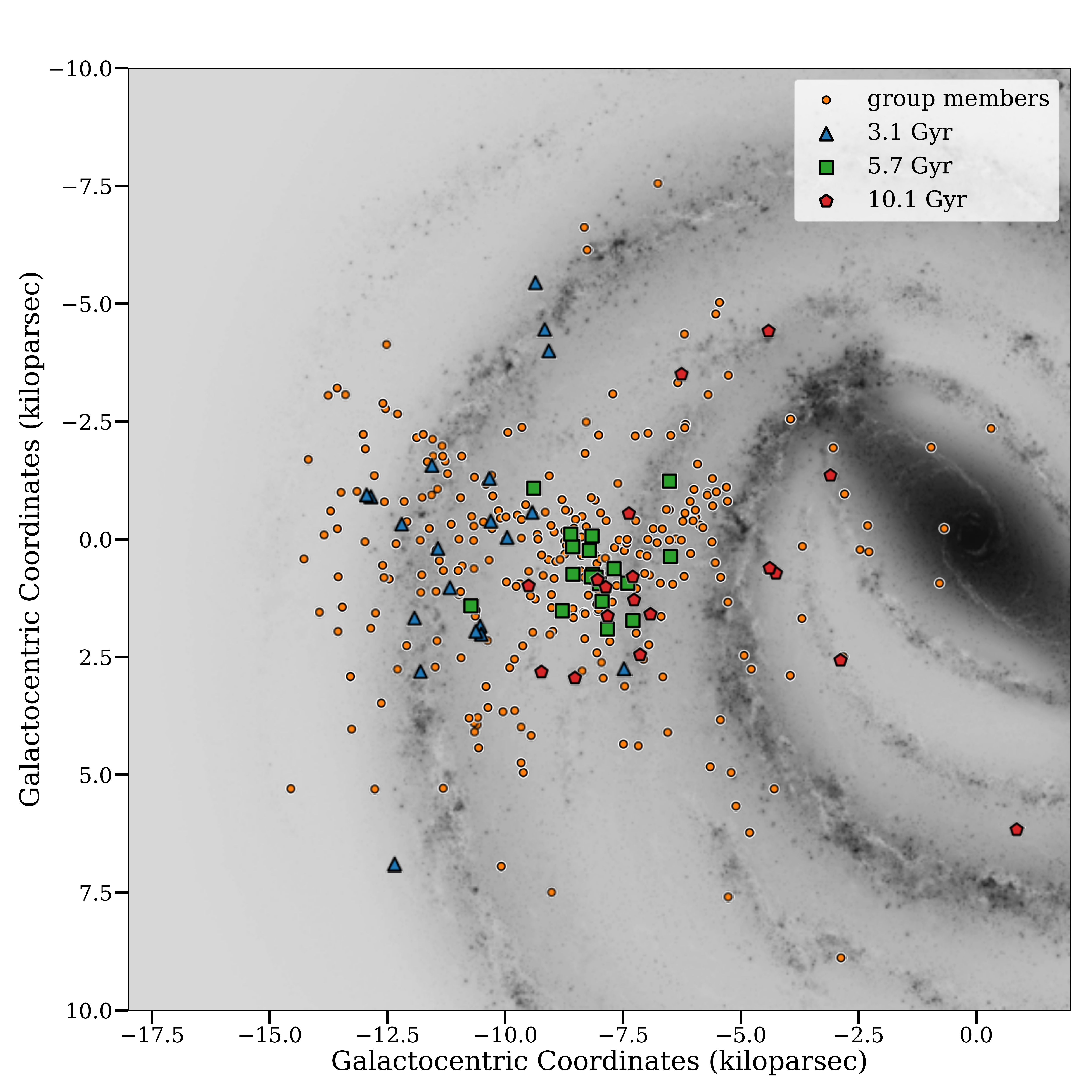}
\caption{Positions of our group members looking down on the Milky Way. Members of our youngest group are shown in blue triangles, median age group members are shown in green squares, and members of the oldest group are shown in red pentagons. Members of the other eighteen groups are shown in orange circles. Members of all groups are widely dispersed across the Galactic disc. The background image is modified from original artist's conception, image credit: NASA/JPL-Caltech/R. Hurt (SSC/Caltech).}
\label{fig:mw}
\end{figure}

With these restrictions on number of members per group and $S^g$, DBSCAN recovers 21 groups from our APOGEE sample for further consideration.

\section{Properties of candidate birth clusters}
\label{sec:clusters}

\subsection{Overall Galactic and chemical distribution}

Members of 18 of the 21 groups identified by DBSCAN are shown in Galactic coordinates in Figure~\ref{fig:onsky} as orange circles. We have highlighted the members of the three remaining groups with different colours and symbols: the group with the oldest median stellar age (10.1 Gyr) is shown as red pentagons, the median age group (5.7 Gyr) is shown using green squares, and the youngest group (3.1 Gyr) is shown with blue triangles. Figure~\ref{fig:onsky} allows us to quickly observe that the members of our groups are distributed across the entire APOGEE sample. The oldest group has the most members at large Galactic latitude, while the youngest group is largely confined close to the Galactic mid-plane. 

An alternate, top-down, perspective on the positions of these group members in physical space is shown in Figure~\ref{fig:mw}, with an artist's conception of the Milky Way used as the background. Members of the same three groups are highlighted, using the same symbols. As shown in Figure~\ref{fig:onsky}, the members of our chemically-tagged groups are broadly distributed in physical space.

\begin{figure}
\centering
\includegraphics[width = \linewidth]{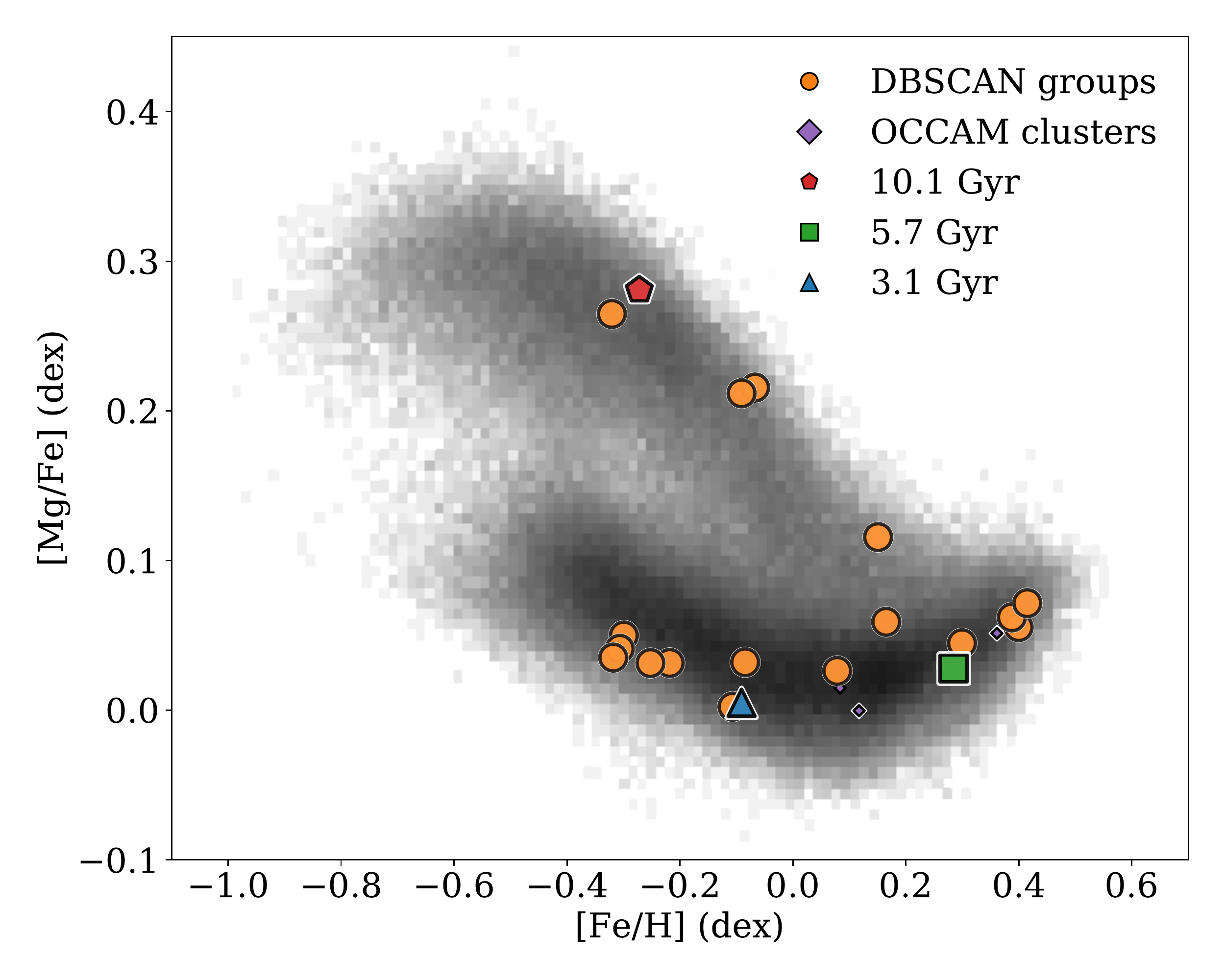}
\caption{The distribution of identified groups in a two-dimensional projection of the chemical space in which we perform chemical tagging. The median abundance positions of the groups are shown with orange circles, with the oldest, median age, and youngest group positions highlighted with a red pentagon, a green square, and a blue triangle, respectively. The distribution of the groups roughly follows the underlying distribution of all APOGEE stars considered for tagging. Shown for comparison in purple diamonds are the median abundance locations of open clusters identified in the OCCAM survey \citep{Donor2020}.} 
\label{fig:mgfe}
\end{figure}

In Figure~\ref{fig:mgfe} we display the positions of our twenty-one groups in a two dimensional projection of our chemical space. Because, as we will see, our groups are very compact in chemical space, the orange circles now mark the positions of the median abundances in [Mg/Fe] and [Fe/H] for each group, instead of the positions of the individual members. For comparison, we also include the median positions of open clusters from the Open Cluster Chemical Abundances and Mapping (OCCAM) survey \citep{Donor2020} that have fifteen or more members in our sample after we make the quality cuts described in \S\ref{sec:chemistry}. We mark the abundance positions of the groups with the oldest, median, and youngest median member age with a red pentagon, a green square, and a blue triangle, respectively. We will continue to consider these three groups as our examples in subsequent analysis.

\begin{figure*}
\centering
\begin{subfigure}{\textwidth}
\includegraphics[width = \linewidth]{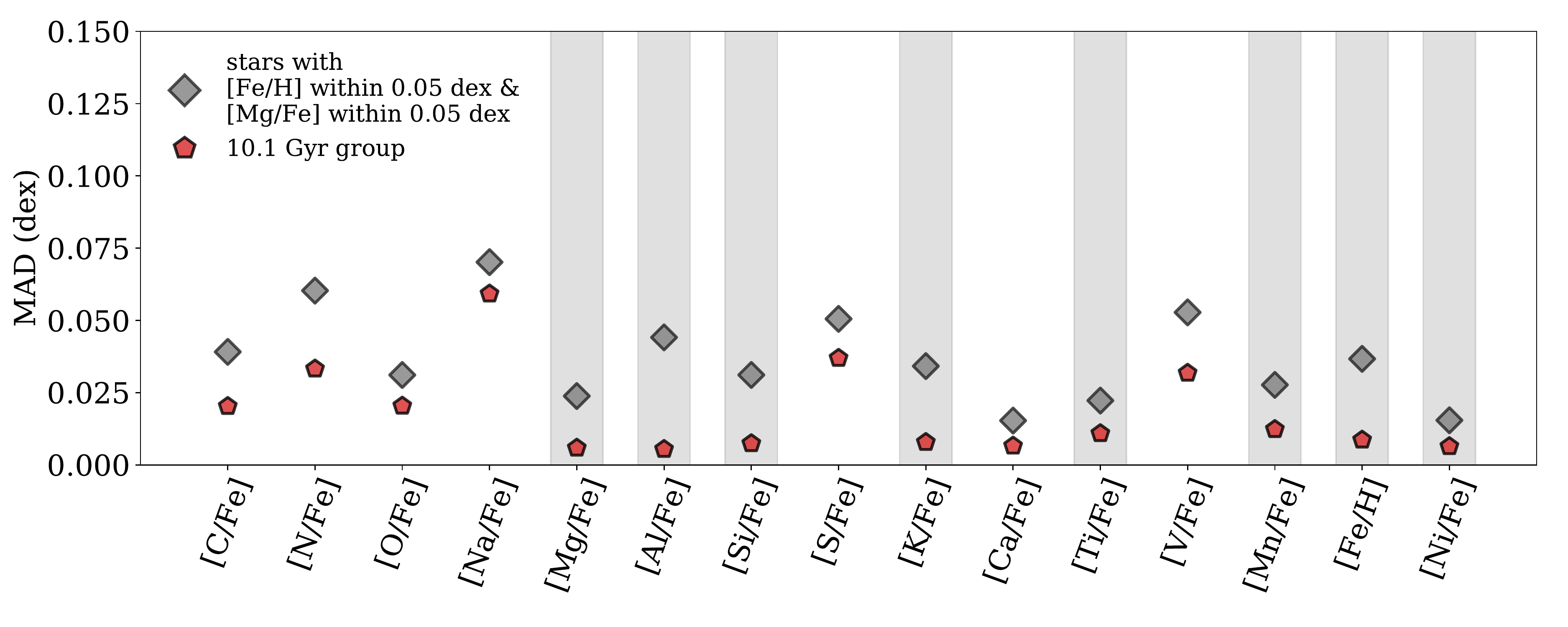}
\label{fig:signatureold}
\end{subfigure}
\newline
\begin{subfigure}{\textwidth}
\includegraphics[width = \linewidth]{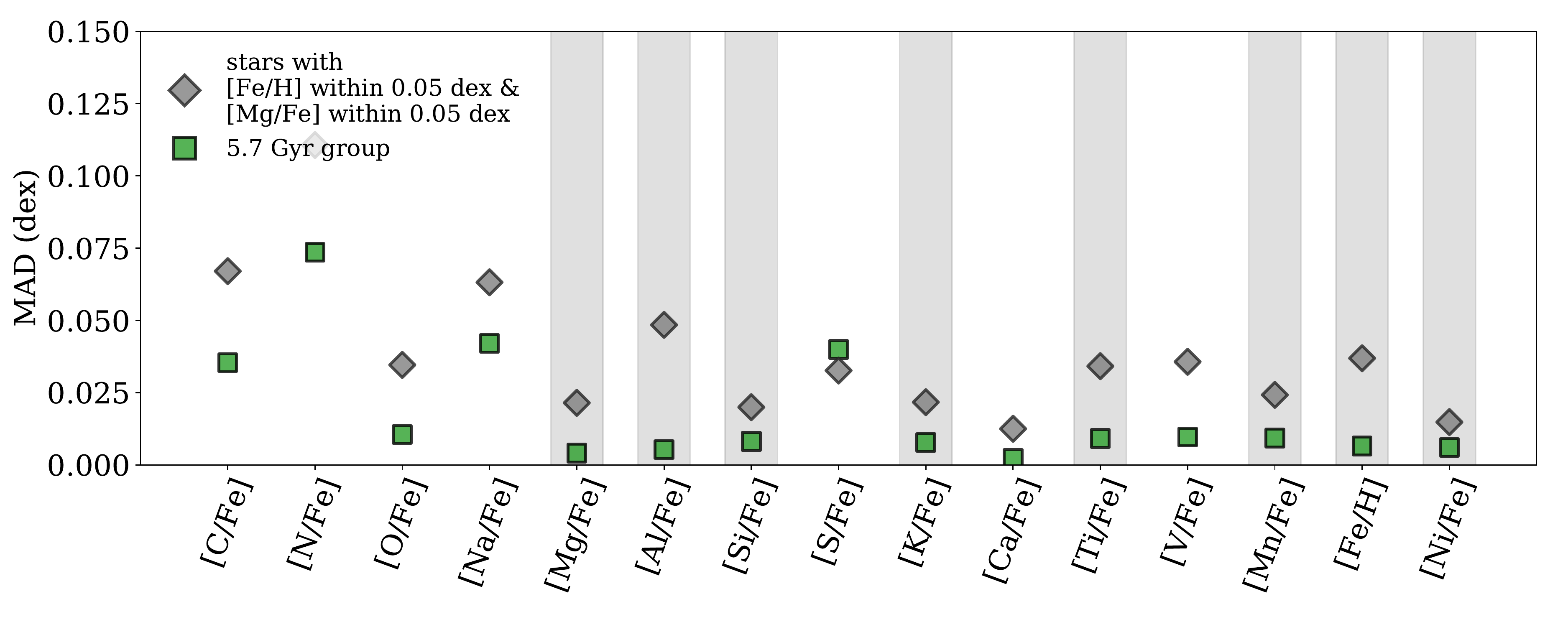}
\label{fig:signaturemed}
\end{subfigure}
\newline
\begin{subfigure}{\textwidth}
\centering
\includegraphics[width = \linewidth]{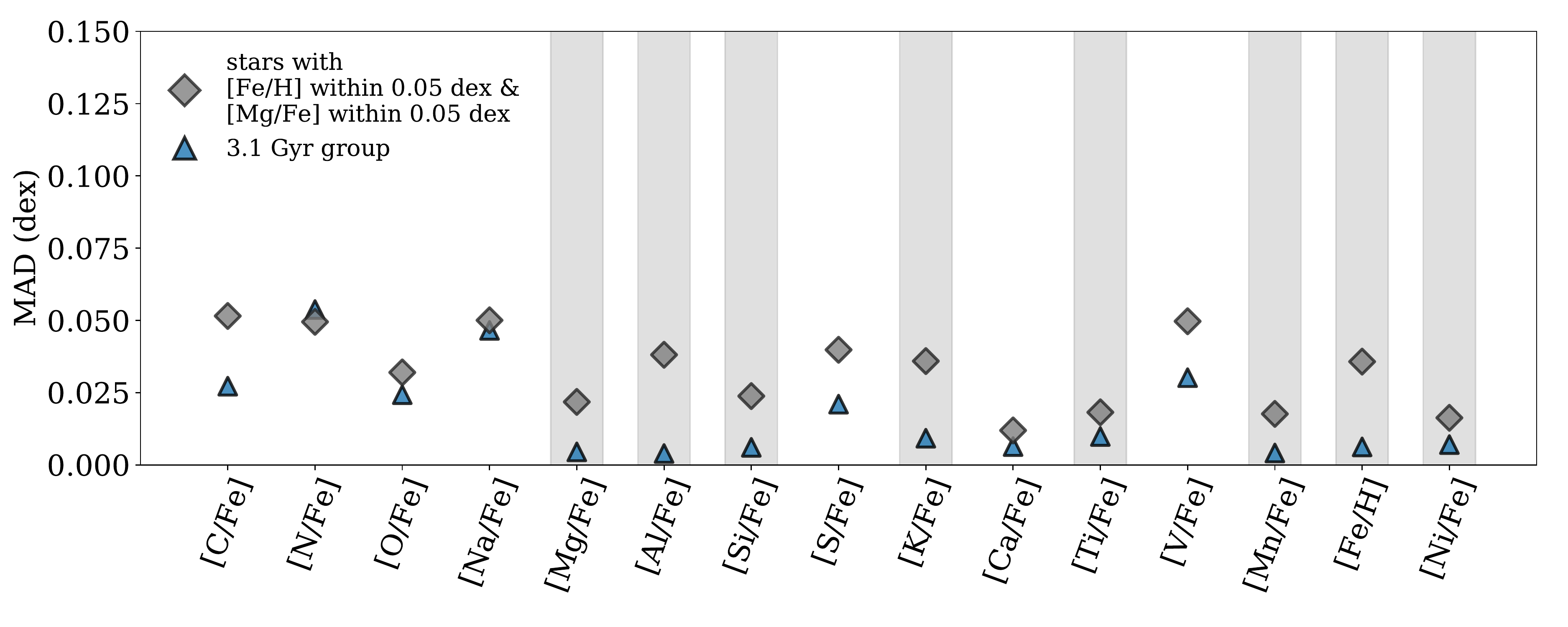}
\label{fig:signatureyoung}
\end{subfigure}
\caption{The MAD of abundances within our three example groups as a function of abundance, shown for our oldest (red pentagons), median age (green squares) and youngest group (blue triangles). For comparison, the MAD for all stars with [Fe/H] and [Mg/Fe] within 0.05 dex of the group median (chemically similar population) are shown as grey diamonds in each subfigure. Abundances used for chemical tagging are highlighted with grey bars. As expected, the abundances used for chemical tagging have the lowest intra-group spread, but our groups are more chemically homogeneous than the chemically similar population in most abundances.}
\label{fig:signature}
\end{figure*}

\begin{figure*}
\centering
\includegraphics[width = \linewidth]{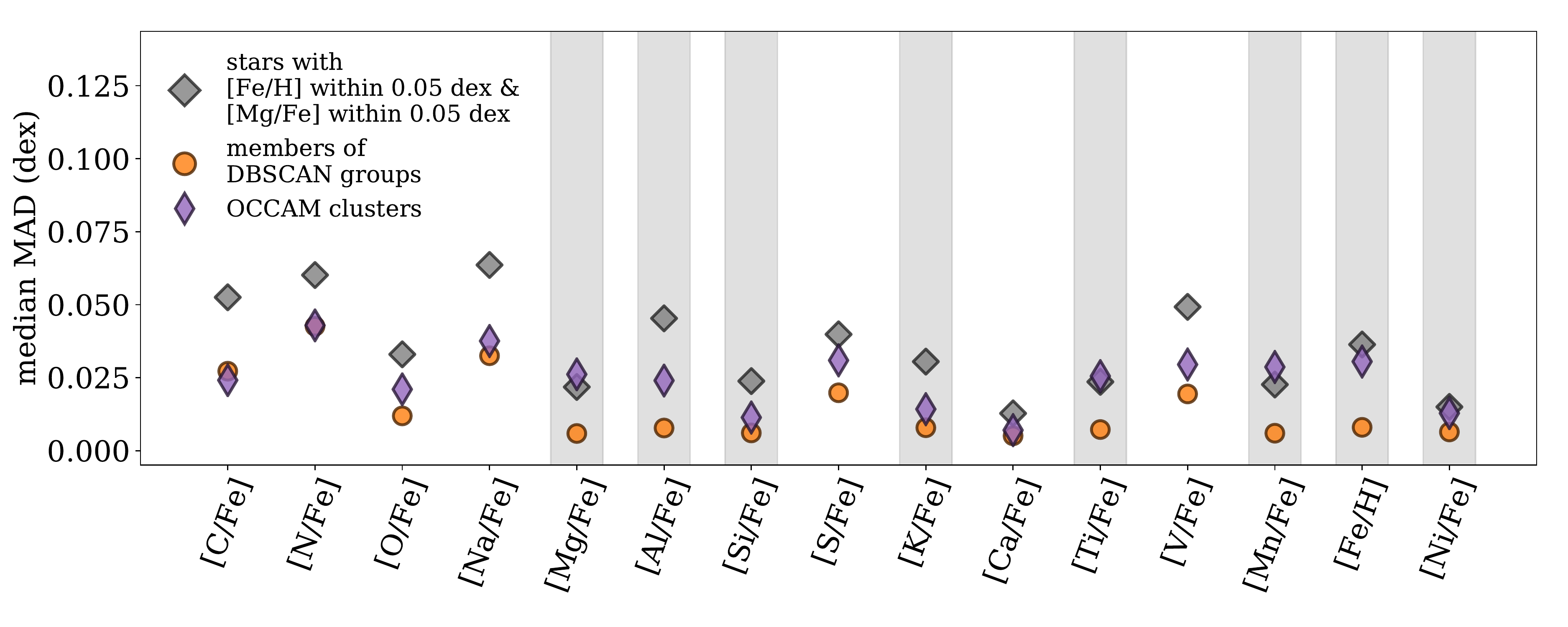}
\caption{This figure takes the median of results like those shown in Figure~\ref{fig:signature}, across all 21 groups. Orange circles show the median value of the abundance spread within a group (measured with the MAD). For comparison, the median MAD values of the stars considered chemically similar to each group are shown with grey diamonds, and the median MAD values for the open clusters in OCCAM are shown with purple diamonds. Our candidate birth clusters are overall more compact than their chemically-similar populations for all elements. They are also more compact that the OCCAM open clusters in most abundance ratios.}
\label{fig:medMAD}
\end{figure*}

\subsection{Chemical homogeneity}

Since we expect birth clusters to be homogeneous in most abundances, not just those we chose for chemical tagging, we consider the spread within our three example groups in each of the 15 elements reliably measured by \texttt{astroNN}. We calculate this spread with the median absolute deviation (MAD). In this work, this and all subsequent calculations of the MAD are multiplied by a factor to make them consistent with comparison to the standard deviation of the distribution; this factor is approximately 1.4826. The median absolute deviation (MAD) of the elemental abundance values within our three example groups are shown in Figure~\ref{fig:signature}, with the elements used for chemical tagging highlighted with grey bars.

The MAD within the group members is compared to the MAD across all stars in our APOGEE sample that have [Fe/H] and [Mg/Fe] within 0.05 dex of the group median value for those abundances (grey diamonds). We hereafter refer to this selection as the `chemically similar' population, and it is different for each group. Selecting these stars allows us to compare our group to a population that came from gas with moderately similar chemical evolution. Our example groups are more chemically homogeneous than the chemically similar population, even in abundances we do not use for chemical tagging. These results are typical for all of our clusters, and demonstrate that with just eight abundances we are able to find groups that are remarkably chemically homogeneous relative to the rest of the APOGEE sample.

Figure~\ref{fig:medMAD} shows the median of the MADs in abundance values across all twenty-one of our groups (orange circles) compared to the median across the MADs of their corresponding chemically similar populations (grey diamonds). This figure demonstrates that all of our groups are typically more homogeneous than their corresponding chemically similar population in all elements. Overall, the groups are most compact relative to their chemically similar population in the abundances we used for chemical tagging, because the spread within the groups for these elements is minimized by our clustering algorithm in order to identify groups. In addition to taking the median, we also employ a simple counting statistic and find that 67\% of our groups were more homogeneous than the chemically similar population in all abundances. All but one of the remaining groups were only less chemically homogeneous than the chemically similar population in a single abundance, and the final group less homogeneous in two (with [N/Fe] and [S/Fe] the worse offenders).  Thus, when considering the spread in  abundance ratios \emph{not} used in our chemical clustering, we find that the detected groups are more homogeneous than chemically-similar, unassociated APOGEE stars.

\subsection{Age homogeneity}

Having satisfied ourselves that our DBSCAN groups are more chemically homogeneous than a chemically-similar population in APOGEE, we consider their ages as derived in \citet{Mackereth2019}. For each group, we find that all members have ages that are the same as the median group age within uncertainties. 

\begin{figure*}
\centering
\begin{subfigure}{0.32\linewidth}
\includegraphics[width = \linewidth]{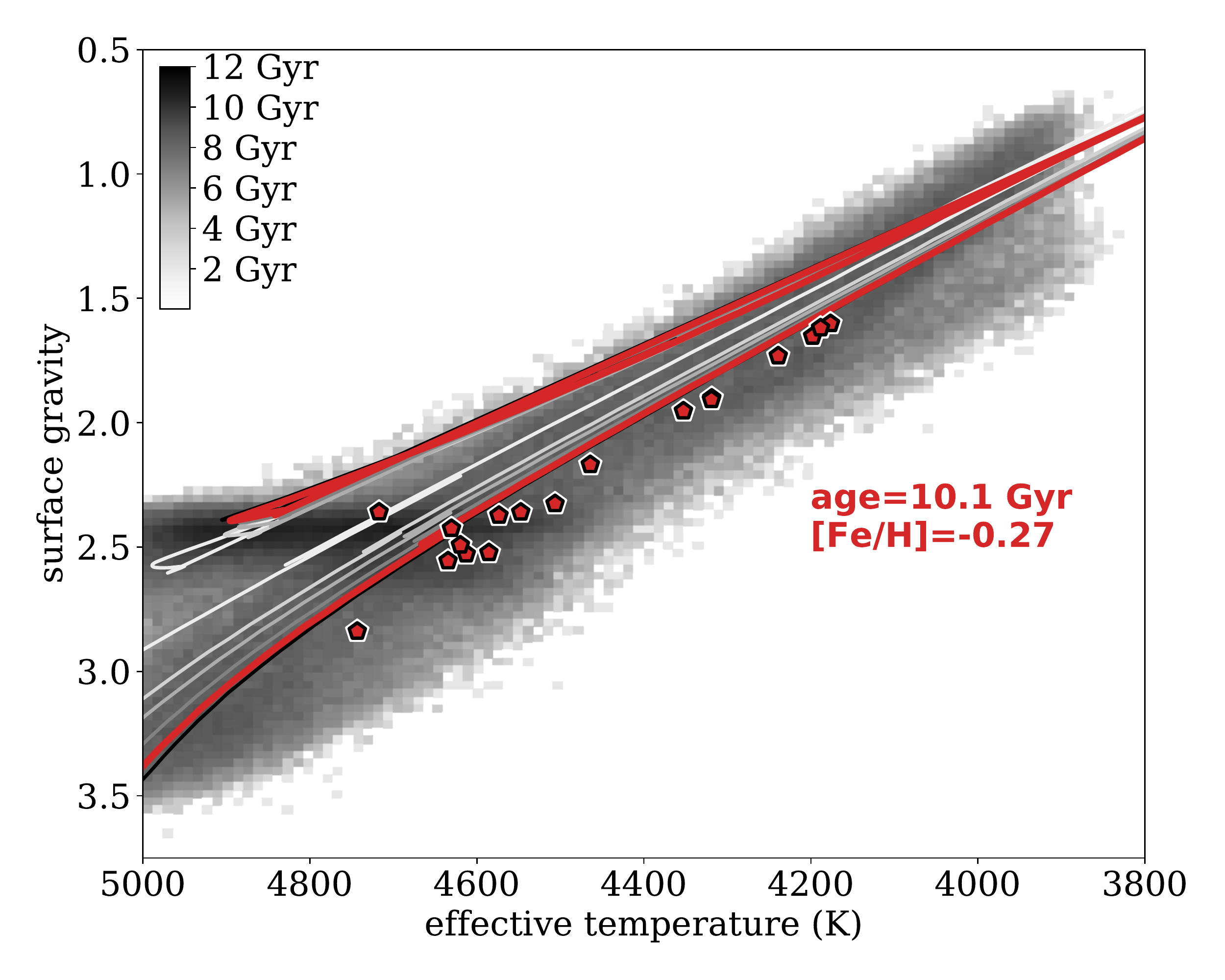}
\end{subfigure}
\begin{subfigure}{0.32\linewidth}
\includegraphics[width = \linewidth]{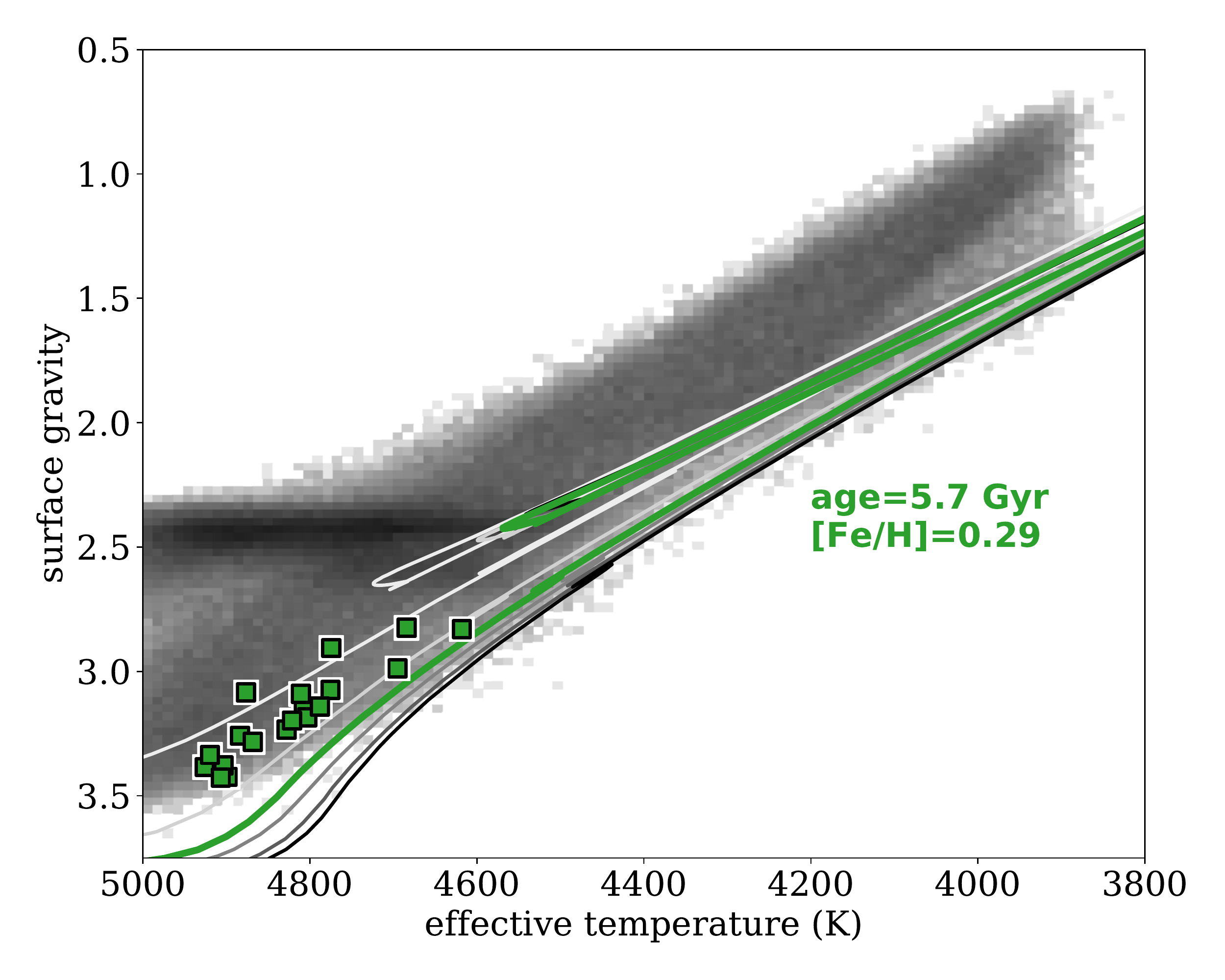}
\end{subfigure}
\begin{subfigure}{0.32\linewidth}
\includegraphics[width = \linewidth]{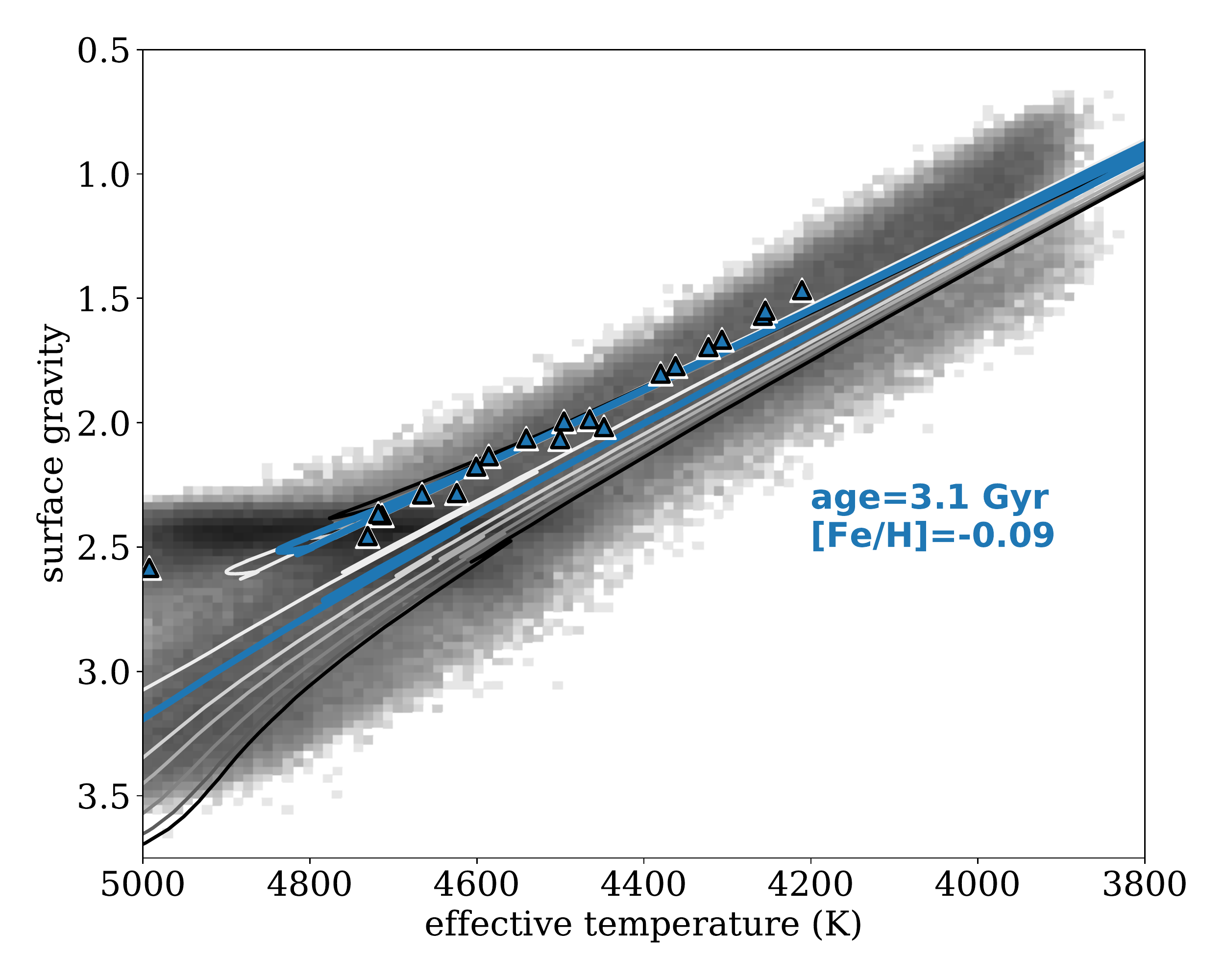}
\end{subfigure}
\caption{Members of our three example groups compared to a PARSEC isochrone \citep{Brennan2012} that most closely matches the median group age and metallicity (thickest line). Members of our oldest group are shown on the left with red pentagons, while members of the median age and youngest groups are shown in green squares and blue triangles in the centre and right panels respectively. For comparison, isochrones with the same metallicity as the group but at ages 2, 4, 6, 8, 10, and 12 Gyr are shown as grey lines (age increases towards the bottom right). The underlying histogram shows the positions of all stars considered for chemical tagging.}
\label{fig:isochrone}
\end{figure*}

Establishing a shared age within these groups is a strong indication that the groups found by DBSCAN are true birth clusters. However, the difficulty of constraining ages for our groups becomes clear when we consider Figure~\ref{fig:isochrone}, in which we show a spectroscopic Hertzsprung-Russell diagram for each of our three example groups, with the APOGEE sample shown as a background histogram. Overplotted in each panel are several isochrones created with the PAdova and TRieste Stellar Evolution Code (PARSEC; \citealt{Brennan2012}). The grey PARSEC isochrones in each panel are separated by 2 Gyr, starting from a 2 Gyr isochrone at the top and proceeding to a 12 Gyr isochrone, and have a metallicity that is the closest match to median group metallicity. Each panel also shows as a thicker line the isochrone closest to the median age of the group members (the largest age difference between median group age and corresponding isochrone age is 0.3 Gyr). All three groups are in reasonable agreement with their isochrones, although the median age group shows the greatest displacement, perhaps due to its higher metallicity. The small separation between the grey reference isochrones highlight the challenge in assigning an accurate age when fitting to stars on the red giant branch. As our investigation of ages is meant to serve only as a consistency check for the plausibility of interpreting our groups as birth clusters, we postpone performing a formal fit for the age of each group using their members' position in the $\log g$ and $\Teff$ to future work. However, the tightness of our groups' $\log g$ vs $\Teff$ relation shows that such a fit will provide a high-precision measurement of their ages.

\section{Orbital diffusion in the Galactic disc}
\label{sec:dyn}

To fully leverage all of the information at our disposal, we make use of the kinematics measured by \emph{Gaia} to compute orbital actions for each star as described in \S\ref{sec:kinematics}. If stars simply orbited in an axisymmetric gravitational potential, their orbital actions would be conserved over their lifetimes. However, the Milky Way's potential is not axisymmetric. Two-body interactions within a birth cluster, as well as tidal heating and stripping will cause a spread in orbital actions among stars born in the same birth cluster as it dissolves, even though we would otherwise expect them to have comparable actions. In addition to these effects, orbital diffusion due to stochastic heating and radial migration causes the actions of stars to drift further over time. Because the members of a birth cluster all start out with very similar actions, their current spread in actions provides a novel and stringent constraint on the cluster dissolution process and the drivers of orbital diffusion in the Galactic disc, especially when coupled with constraints on the chemical evolution of the disc. In this section, we perform a first, cursory investigation of how a birth cluster's orbital actions change as it evolves in a realistic Galactic potential, but we defer more extensive modeling of the effect of different drivers of diffusion to future work. However, despite the many mechanisms that might evolve the orbital actions of a star over its lifetime, we still expect some upper limit on the possible range of actions displayed by members of the same birth cluster.

\begin{figure*}
\centering
\includegraphics[width = \linewidth]{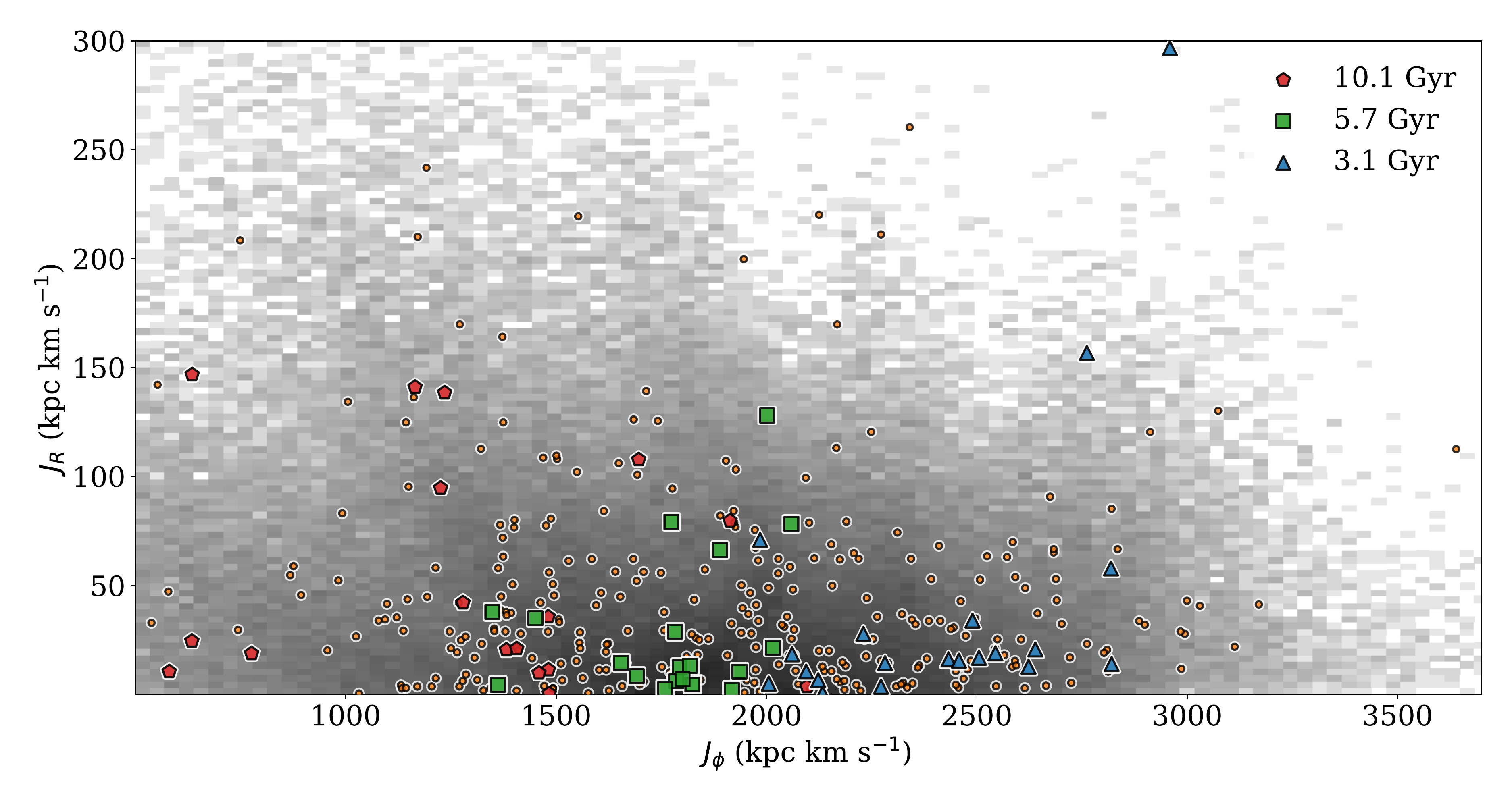}
\caption{The distribution of group members in action space (orange circles), with the youngest, median age, and oldest group members highlighted as blue triangles, green squares, and red pentagons, respectively. As our groups increase in age, the azimuthal actions of their member stars increases.}
\label{fig:actions386}
\end{figure*}

The distribution of $J_{\rm R}$ vs $J_{\phi}$ for all stars considered for chemical tagging is shown as as a two-dimensional histogram in Figure~\ref{fig:actions386}. Plotted over that distribution are the actions of all stars identified as group members (orange circles) with those of our oldest, median, and youngest age groups highlighted in red pentagons, green squares, and blue triangles, respectively. To assess whether our groups are kinematically consistent with being dispersed birth clusters, we turn to $N$-body simulations of star cluster evolution, with the aim of reproducing the spread in the actions observed in each group. The action spread varies dramatically between groups, and we characterize it with the MAD of the actions of the group members. 

Our simulation uses the Barnes-Hut Tree code \citep{Barnes1986} implemented in Astrophysical Multipurpose Software Environment (\texttt{AMUSE}; \citealt{Portegies2013, Pelupessy2013, Portegies2018}) to resolve intra-cluster dynamics. The setup is heavily inspired by \citet{Webb2019}, and while we summarize it here, interested readers may refer to that work for a detailed explanation of our set up. 

To evolve our birth clusters, we consider them as $N$-body systems within a larger galaxy, which we describe with a series of increasingly complex Galactic potentials created with \texttt{galpy}. We begin with a static Galactic potential, choosing to use \texttt{MWPotential2014}, the components of which describe the Galactic halo, disc, and bulge \citep{Bovy2015a}. We refer to this as our `static' case.

In addition to the static potential case we consider three time-varying potentials, each of which incorporates a Galactic bar from \citet{Dehnen2000}, with bar length of 5 kpc and a pattern speed $35.75\,\mathrm{km} \mathrm{s}^{-1} \mathrm{kpc}^{-1}$ \citep{Portail2017, Sanders2019, Bovy2019}. All three time-varying cases also include two-armed  spirals, modelled with the sinusoidal potential from \citet{Cox2002} and implemented in \texttt{galpy} by \citet{Hunt2018}. All of the \texttt{galpy} potentials, including the bar and spiral arm potentials, can be used in AMUSE using the glue code described in \citet{Webb2019}.

The spiral arms have the same pattern speed as the bar, a density of 0.13 $M_{\odot} \mathrm{pc}^{-3}$, and a pitch angle of $25^{\circ}$. In our `bar + arms' case, we model the spiral arms as a density wave with the same pattern speed as the bar. In our `bar + transient arms' case, the amplitude of the spiral arm potentials varies with time and is modulated with a Gaussian, as discussed in detail in \citet{Hunt2018}. Each pair of spiral arms persists for a 460 Myr episode before fading away. At any time, there are three sets of two spiral arms extant in the potential: one pair with its amplitude growing, one dominant pair with the largest amplitude, and one pair with its amplitude fading. In our final case, the `bar + transient winding arms', the spiral arms retain their transient nature and are additionally everywhere co-rotating, winding up over time \citep{Grand2012}.

For each of our four options for the Galactic potential, we consider each group found by DBSCAN and generate a suite of simulations. The orbit of each star in the group is reverse integrated with \texttt{galpy} in each potential from its current phase space position for a time equal to the median age of the group. Assuming valid age estimates, the position of the star after this integration should be where its birth cluster was born. We use this phase space position as the initial conditions for our $N$-body simulation. Each cluster is initialized according to a Plummer model with a half-mass radius of 3 pc and a mass of 1000 $M_{\odot}$ distributed equally to 1000 members, centred at the chosen initial conditions. We use a softening length of 3 pc and an opening angle of 0.6 radians. With these choices, our simulated birth clusters disrupt fairly quickly: since our groups do not appear to form streams, we know their birth clusters should be fully dispersed. We evolve the simulated birth cluster forward in our potential for a time equal to the median age of the group using 1 Myr timesteps. Once this evolution is complete, we compute the radial and azimuthal actions for each of the $N$-body stars using the St\"ackel approximation \citep{Binney2012}, assuming the potential is \texttt{MWPotential2014}. We then calculate the MAD of $J_R$ and $J_{\phi}$ across all 1000 of our simulated stars. Once we have generated a simulation for every group member in a given group, we take the median of the MADs of both $J_R$ and $J_{\phi}$ for each simulation to compare with the observed MAD for those actions. In this way, we make a measurement of the typical range in actions expected for stars that were born in the same birth cluster. We expect this spread in actions to increase with cluster age, since older clusters will have had longer to diffuse through phase space.

\begin{figure}
\centering
\includegraphics[width = \linewidth]{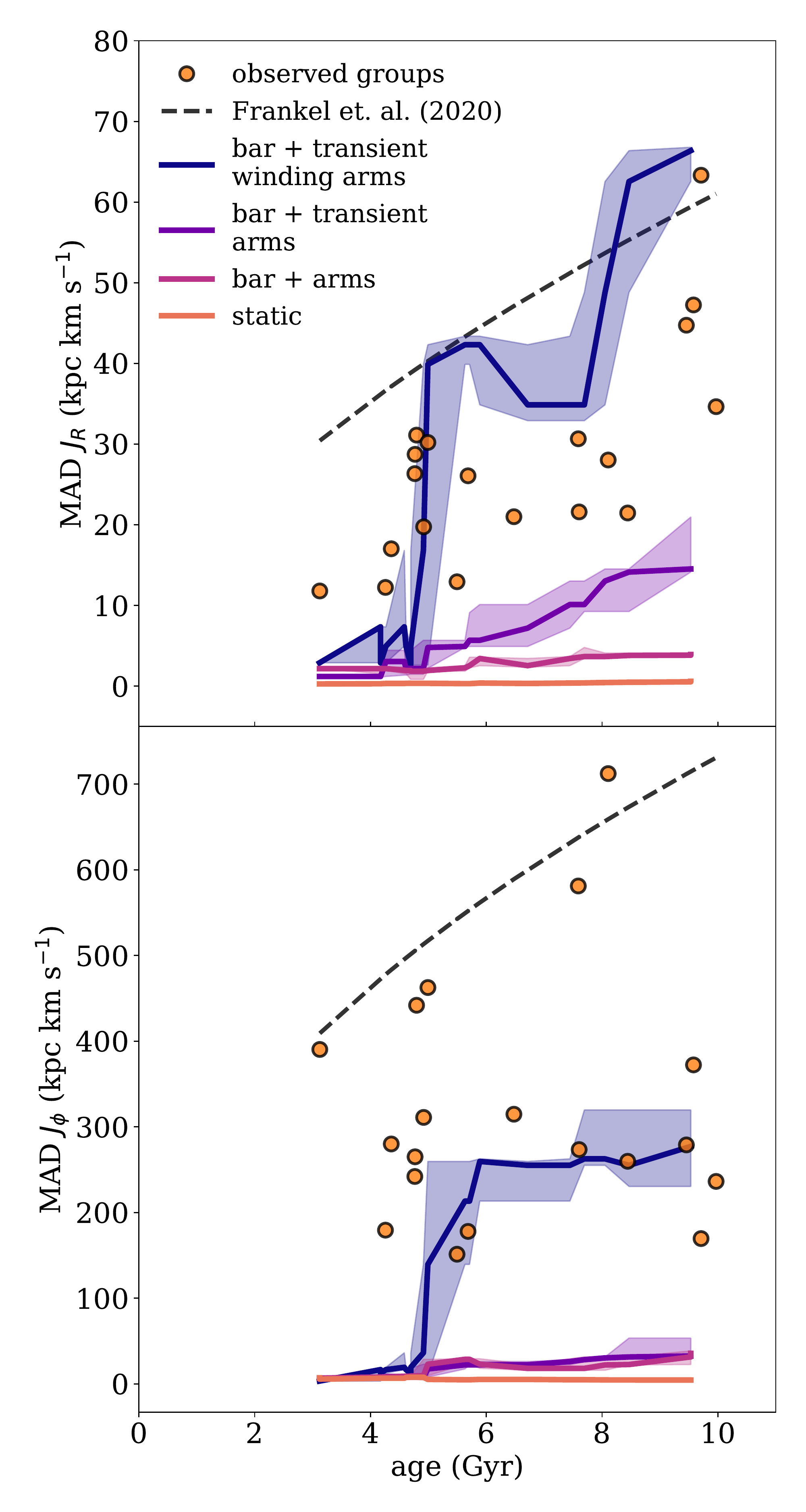}
\caption{The scaled median absolute deviation (or spread) of group member actions as a function of median group member age. The solid line shows the rolling median the action spread of our simulated clusters at the same age. Shaded regions show the interquartile range in the spread. The grey dashed lines show expected deviation from birth action as a function of star age from \citet{Frankel2020}}
\label{fig:actions}
\end{figure}

The resulting action MAD for each of our three potential cases are shown as a function of group age in Figure~\ref{fig:actions}. We have taken a running median of our actions as a function of age to facilitate comparison. In addition, we include the relationship derived in \citet{Frankel2020} for present day action spread as a grey dashed line. In that  work, the authors modelled the change in orbital actions over a star's lifetime due to angular momentum diffusion (which changes $J_{\phi}$) and radial heating (which changes $J_R$). Their model predicts the change $J_R$ and $J_{\phi}$ as a function of age (and Galactocentric radius, in the case of $J_R$), assuming that co-eval stars are born on circular orbits with [Fe/H] unique to the time of birth and the initial angular momentum. We have used their relations for present day spread, assuming a Galactocentric radius of 8 kpc for the $J_R$ relation. \citet{Frankel2020} fit their model to red clump stars from APOGEE, and generally predict greater changes in orbital actions than we observe in our $N$-body simulations, but a key difference is that they constrain diffusion of stars born at the same time and at the same Galactocentric radius, but not the same Galactocentric azimuth, while we include the latter. Comparing the observed present-day action spread for our candidate birth clusters with the heating and migration behaviour found by \citet{Frankel2020} based on fitting the chemical properties of stars born in annuli across the disc, we see that we observe much less action spread as a function of age, although the ratio in the spread in $J_R$ and $J_\phi$ is similar to their ratio. That we see less action spread than required in Frankel et al. (2020)'s ring-to-ring diffusion model indicates that ring-to-ring diffusion has a significant contribution from the stochasticity of where in the ring stars start out.

Consistent with our expectations, the spread in $J_R$ tends to increase as a function of group age, with all four of our potential cases exhibiting the same trend.  Simulations in a `static' potential dramatically under-predict the spread in $J_{\phi}$. Our `bar + arms' and `bar + transient arms' cases come closer, but still under-predict, while our `bar + transient winding arms' case slightly over-predicts. In both data and in simulations, there exists much less of an age trend for the spread in $J_{\phi}$ than the spread in $J_R$. All four of our potential cases under-predict the spread in $J_{\phi}$, although our `bar + transient winding arms' case comes the closest. 

Comparing the large spread in the actions for a given group as a function of age with the small spread predicted for the static, axisymmetric potential clearly demonstrates that we need substantial orbital heating (in $J_R$) and migration (in $J_\phi$) to explain the observed action spread. Heating and migration caused by transient spiral structure can qualitatively explain the observed trends of action spread with age. But to match them in detail will require a more extensive set of simulations that vary the properties of the spiral-structure episodes and that also consider the effect of diffusion driven by encounters with GMCs and by satellite heating. Adding these mechanisms for diffusion tends to increase the range of available actions, as seen in \citet{Webb2019, Jorgensen2020}, and thus may more accurately reproduce our the action spread we observe for our groups.

\section{Discussion}
\label{sec:discussion}

The groups we have identified with DBSCAN in this work exhibit many properties we would predict for birth clusters, given their homogeneous abundances and ages. To that end we classify them as birth cluster candidates. A full catalogue of our candidate clusters and their membership is available online \footnote{\url{https://doi.org/10.5281/zenodo.3909859}}. We summarize the properties of our candidate clusters in Table~\ref{tab:candidate}.
\begin{table*}
\caption{The median properties of our candidate birth clusters. Median cluster ages are accompanied by the median absolute deviation of ages within the cluster in brackets.}
\label{tab:candidate}
\centering
	\begin{tabular}{r|c|c|c|c|c|c|c|c|c|c}
		ID & [Fe/H] & [Mg/Fe] & [Al/Fe] & [Si/Fe] & [K/Fe] & [Ti/Fe]  & [Mn/Fe] & [Ni/Fe] & age in Gyr & \# of members  \\
		\hline
        PJ01 & -0.32  & 0.03 & -0.06 & 0.05  & 0.00  & -0.10 & -0.08 & 0.03 & 4.7 (0.4) & 16\\
        \rowcolor{Gray}
        PJ02 & -0.32  & 0.26 & 0.29 & 0.15  & 0.21  & -0.00 & -0.15 & 0.09 & 9.5 (0.6) & 15\\
        PJ03 & -0.31  & 0.04 & -0.05 & 0.06  & -0.00  & -0.10 & -0.09 & 0.02 & 4.6 (0.6) & 16\\
        \rowcolor{Gray}
        PJ04 & -0.30  & 0.05 & 0.03 & 0.08  & -0.00  & -0.10 & -0.09 & 0.03 & 4.9 (0.5) & 15\\
        PJ05 & -0.27  & 0.28 & 0.30 & 0.17  & 0.19  & -0.00 & -0.16 & 0.09 & 10.1 (0.4) & 18\\
        \rowcolor{Gray}
        PJ06 & -0.25  & 0.03 & -0.01 & 0.04  & 0.00  & -0.10 & -0.08 & 0.02 & 4.7 (0.4) & 16\\
        PJ07 & -0.21  & 0.03 & -0.07 & 0.04  & 0.01  & -0.07 & -0.07 & 0.01 & 4.6 (0.3) & 19\\
        \rowcolor{Gray}
        PJ08 & -0.11  & 0.00 & -0.08 & 0.02  & -0.02  & -0.03 & -0.06 & -0.01 & 4.3 (0.6) & 17\\
        PJ09 & -0.09  & 0.21 & 0.25 & 0.10  & 0.14  & 0.01 & -0.11 & 0.05 & 9.5 (0.5) & 19\\
        \rowcolor{Gray}
        PJ10 & -0.09  & 0.00 & -0.00 & 0.03  & -0.04  & -0.08 & -0.05 & -0.02 & 3.1 (0.5) & 20\\
        PJ11 & -0.08  & 0.03 & 0.03 & 0.01  & 0.00  & -0.01 & -0.05 & 0.02 & 4.2 (0.9) & 16\\
        \rowcolor{Gray}
        PJ12 & -0.07  & 0.22 & 0.27 & 0.10  & 0.14  & 0.04 & -0.11 & 0.05 & 9.7 (0.1) & 16\\
        PJ13 & 0.08  & 0.03 & 0.02 & 0.01  & -0.01  & 0.06 & -0.02 & 0.01 & 6.7 (0.4) & 15\\
        \rowcolor{Gray}
        PJ14 & 0.15  & 0.12 & 0.17 & 0.02  & 0.05  & 0.05 & -0.01 & 0.03 & 8.5 (0.3) & 15\\
        PJ15 & 0.17  & 0.06 & 0.09 & -0.02  & 0.03  & 0.06 & 0.02 & 0.03 & 7.4 (0.6) & 15\\
        \rowcolor{Gray}
        PJ16 & 0.28  & 0.03 & 0.03 & -0.01  & -0.00  & 0.06 & 0.08 & 0.03 & 5.0 (1.0) & 17\\
        PJ17 & 0.29  & 0.03 & 0.06 & -0.04  & 0.01  & 0.09 & 0.09 & 0.06 & 5.7 (1.0) & 19\\
        \rowcolor{Gray}
        PJ18 & 0.30  & 0.05 & 0.09 & 0.00  & 0.00  & 0.13 & 0.06 & 0.03 & 7.7 (0.3) & 17\\
        PJ19 & 0.39  & 0.06 & 0.07 & -0.00  & 0.01  & 0.11 & 0.12 & 0.03 & 5.6 (0.5) & 19\\
        \rowcolor{Gray}
        PJ20 & 0.40  & 0.05 & 0.13 & -0.03  & 0.03  & 0.12 & 0.12 & 0.08 & 5.9 (0.7) & 24\\
        PJ21 & 0.41  & 0.07 & 0.10 & 0.01  & 0.02  & 0.19 & 0.10 & 0.05 & 8.1 (0.5) & 16\\
	
	\end{tabular}
\end{table*}

\subsection{Comparison to groups from simulated chemical space}

To investigate the validity of our DBSCAN-identified groups as birth cluster candidates, we analyze the groups found by DBSCAN in the eight-dimensional simulated chemical space described in \S\ref{sec:sim}. Using our usual criteria of maximizing the number of groups recovered, we find that DBSCAN prefers a larger $\epsilon$ value (i.e. chemical neighbourhood size) for the simulated chemical space (0.08 dex, vs 0.02 dex for the observed abundances). We identify 67 groups in the simulated chemical space, up from the 21 found in the observed abundances.

For each of the 67 groups identified in simulated chemical space, we compute the group homogeneity. We do this by looking at the true cluster labels for the stars in each of our groups. The true cluster that contributes the most member to the group is considered the match; the homogeneity is the fractional membership of this matched cluster. Therefore if our found group has twenty members and we see that ten come from one true cluster, five from another, and five from a third, the homogeneity score for that group would be $10/20=0.5$. The groups we find in our simulated chemical space have a median homogeneity of 0.94, with a median absolute deviation of 0.09.

We also compute group `completeness', which takes the matched true cluster for each group and calculates the fraction of members of that true cluster that ended up in the group. This completeness score is typically lower for our procedure, as we select parameters for DBSCAN to prefer higher homogeneity at the expense of completeness. The two scores tend to have a reciprocal relationship; increasing one generally lowers the other, except in the case of perfect clustering. This relation holds true for our simulated chemical space; although we have high homogeneity, the median completeness is 0.24, with a median absolute deviation of 0.11. Our preference for the higher homogeneity allows us to feel confident that while our birth cluster candidates may not include all cluster members, they can still represent the most densely chemically-clustered parts (chemical space cores) of dispersed birth clusters.

\subsection{Comparison to open clusters}

As described in \S\ref{sec:intro}, open clusters are often considered proxies for birth clusters, and thus comparison with open clusters offers an additional way to validate our birth cluster candidates. We show in Figure~\ref{fig:mgfe} the positions of the open clusters identified in OCCAM with fifteen or more members in the APOGEE sample considered for chemical tagging. However, as Figure~\ref{fig:mgfe} shows, we do not recover any of these open clusters as birth cluster candidates. This is not as surprising as it might first appear; we find that the MAD of abundances within an open cluster is typically higher than the groups identified by DBSCAN (although open cluster abundance MADs are still usually less than the chemical similar population - see Figure~\ref{fig:medMAD}). This higher internal abundance spread prevents them from being picked out by our algorithm as distinct groups. The open clusters also tend to sit over the densest part of the distribution of APOGEE stars in [Mg/Fe] vs [Fe/H] space, so it is possible that DBSCAN identified them as part of a larger group that did not meet out requirements on silhouette coefficient. 

Open clusters being less chemically homogeneous than our birth cluster candidates does not invalidate them as birth cluster proxies. When applying DBSCAN to our simulated chemical space in \citet{Price-Jones2019}, we showed that DBSCAN often picks out the dense central concentration (or core) of the input clusters, since we optimize for membership homogeneity over completeness. The open cluster members observed in APOGEE likely do not sufficiently sample a dense core that DBSCAN can find when applied to the observed abundances. However, this does not mean that no such core exists. Future chemical tagging experiments should continue to include known clusters to test techniques.

\subsection{Future work}

Our birth cluster candidates are a promising result for chemical tagging. Their shared properties in chemistry and age seem to strongly indicate that they come from the same GMC. From Table~\ref{tab:props}, it can be seen that some of our candidates share similar chemical signatures, sometimes agreeing below the level of our maximum abundance uncertainty cut (0.15 dex). These similar signatures may seem to suggest that some groups  represent separated parts of the same birth associations of stars rather than distinct formation sites. However, many of the abundance differences, while less than the maximum possible abundance uncertainty, are greater than the median abundance uncertainty for our sample ($\leq 0.05$ dex for each abundance considered for chemical tagging). In addition, while differences in an individual abundance may be small, the groups are still quite distinct when distances between them are measured in eight-dimensional chemical space.

Despite their internal chemical homogeneity, it is still possible that our candidates were never gravitationally bound. However, the expected homogeneity of GMCs, as established in simulations like that of \citet{Feng2014}, indicates that at the very least members of each candidate likely came from the same star forming region. \citet{Bland-Hawthorn2010} find that birth clusters with masses up to $10^5 M_{\odot}$ should be chemically homogeneous, but more massive sites may begin to self-pollute, evolving the chemical signature. Accounting for this self-pollution may allow us to identify groups to reconstruct larger birth clusters. If our groups are neither parts of dissolved birth clusters nor the result of a shared formation site, we are left with the question: what chemical evolution history of our Galaxy manufactures such chemical similarity in otherwise unassociated stars, currently spread out over the disc of the Galaxy? Perhaps with larger spectroscopic surveys our groups will be shown to be part of much larger stellar associations, their appearance here merely a product of the APOGEE selection function. If our groups remain chemically distinct but still cannot be validated as birth clusters, they tell us about some consistency across many star formation sites across the Galaxy. This interpretation of our groups as part of larger star formation events would be consistent with \citet{Ness2019}, wherein the authors find that age and [Fe/H] can predict orbital properties and other APOGEE abundance ratios, thus eliminating chemical tagging as a possibility in current surveys. However, our results here demonstrate it is still possible to find chemically homogeneous groups that would not be identified with [Fe/H] and age alone; our results for Figure~\ref{fig:signature} look nearly identical when the chemically similar population is further reduced to only stars with similar ages to the group.

To truly determine the significance of our groups, we would find it most valuable to include additional abundance measurements, particularly for elements that probe additional nucleosynthetic pathways. In this study, we have chosen the abundances measured by APOGEE that track $\alpha$-, odd-Z and Fe-peak elements. However, including elements like Eu and La to investigate the contribution of the $r$-process, and Ba and Y to explore the significance of $s$-process would greatly strengthen any claim that our candidates truly represent birth clusters. Some elements from these processes are starting to be measured in APOGEE (in particular Nd and Ce, see \citealt{Hasselquist2016, Cunha2017}, respectively), but others are difficult or impossible to determine in the $H$-band. One important continuation of this work will be to follow up the members of our candidates in other wavelengths. This follow-up will have the dual effect of increasing the number of available abundances for chemical homogeneity checks while also confirming that our stars remain clustered when abundances are measured with different lines. In addition, dedicated follow up will be able to discern abundances at higher precision than the automated generation performed by APOGEE, further increasing our certainty that these groups are indeed chemically homogeneous.

\section{Conclusions}
\label{sec:conclusions}
By applying a density-based group finding algorithm to a subset of the \texttt{astroNN} abundances derived from the APOGEE spectroscopic dataset we are able to chemically tag 21 groups of stars with more than 15 members. These groups are chemically homogeneous not just in the abundances we chose to use for chemical tagging but in the other seven abundances well measured by \texttt{astroNN}. Our chosen eight abundances ([Mg/Fe], [Al/Fe], [Si/Fe], [K/Fe], [Ti/Fe], [Mn/Fe], [Fe/H], [Ni/H]) represent the the best measured abundances for the nucleosynthetic pathways probed by APOGEE's automatic abundance generation pipeline.

Our groups of stars are consistent with having the same age (estimated in \citealt{Mackereth2019}), and this age consistency, along with their chemical homogeneity makes accepting them as stellar birth clusters a tantalizing prospect. To further investigate the possibility of shared origins for the stars in our groups, we use \emph{Gaia} parallax, position, and proper motion, along with APOGEE radial velocities to compute orbital actions ($J_R, J_{\phi}$) for our stars.  We then perform $N$-body cluster simulations with a realistic Galactic potential that includes a time-varying bar and transient winding spiral arms and measure how our simulated cluster disperses in action space. We find that our simulated clusters exhibit action spreads qualitatively similar to the spread in $J_R$ and $J_{\phi}$ observed in our groups. 

Following up our group members at other wavelengths will be of crucial importance to further establishing their potential shared origins. However we have demonstrated in this work the power of a density-based approach to strong chemical tagging, in particular its ability to identify important areas of interest in high-dimensional chemical spaces populated by large numbers of stars. 

If our groups are in fact true birth clusters, their identification offers an unprecedented opportunity to study the star formation history of the Milky Way across nearly the full range of its life. Our groups range in age from 3 to 10 Gyr, allowing us to access star formation at times not typically probed by studying the origins of open clusters. In addition, extensions to our preliminary $N$-body simulations can offer deeper insight into the exact timeframe for the dispersal of these clusters, which in turn can constrain the broader evolution of the Galactic potential over time.

\section*{Data availability}

A catalogue of the groups identified by this study is available online \url{https://doi.org/10.5281/zenodo.3909859}. Code to produce most of the plots in this work is also available online at \url{https://github.com/npricejones/apogeeCT}.

\section*{Acknowledgements}

The authors thank the anonymous reviewer for their insightful feedback on this paper. We also thank Joss Bland-Hawthorn, Jos\'e Fern\'andez Trincado, Neige Frankel, Ken Freeman, Sten Hasselquist, and the APOGEE team for helpful comments and conversations that enhanced this work.

NPJ is supported by an Alexander Graham Bell Canada Graduate Scholarship-Doctoral from the Natural Sciences and Engineering Research Council of Canada. NPJ and JB received support from the Natural Sciences and Engineering Research Council of Canada (NSERC; funding reference number RGPIN-2015-05235) and from an Ontario Early Researcher Award (ER16-12-061). 

The University of Toronto operates on the traditional land of the Huron-Wendat, the Seneca, and most recently, the Mississaugas of the Credit River; the authors are grateful to have the opportunity to work on this land.

The Dunlap Institute is funded through an endowment established by the David Dunlap family and the University of Toronto.

Funding for the Sloan Digital Sky Survey IV has been provided by the Alfred P. Sloan Foundation, the U.S. Department of Energy Office of Science, and the Participating Institutions. SDSS-IV acknowledges support and resources from the Center for High-Performance Computing at the University of Utah. The SDSS web site is \url{www.sdss.org}.

SDSS-IV is managed by the Astrophysical Research Consortium for the 
Participating Institutions of the SDSS Collaboration including the 
Brazilian Participation Group, the Carnegie Institution for Science, 
Carnegie Mellon University, the Chilean Participation Group, the French Participation Group, Harvard-Smithsonian Center for Astrophysics, 
Instituto de Astrof\'isica de Canarias, The Johns Hopkins University, Kavli Institute for the Physics and Mathematics of the Universe (IPMU) / 
University of Tokyo, the Korean Participation Group, Lawrence Berkeley National Laboratory, 
Leibniz Institut f\"ur Astrophysik Potsdam (AIP),  
Max-Planck-Institut f\"ur Astronomie (MPIA Heidelberg), 
Max-Planck-Institut f\"ur Astrophysik (MPA Garching), 
Max-Planck-Institut f\"ur Extraterrestrische Physik (MPE), 
National Astronomical Observatories of China, New Mexico State University, 
New York University, University of Notre Dame, 
Observat\'ario Nacional / MCTI, The Ohio State University, 
Pennsylvania State University, Shanghai Astronomical Observatory, 
United Kingdom Participation Group,
Universidad Nacional Aut\'onoma de M\'exico, University of Arizona, 
University of Colorado Boulder, University of Oxford, University of Portsmouth, 
University of Utah, University of Virginia, University of Washington, University of Wisconsin, 
Vanderbilt University, and Yale University.

This work has made use of data from the European Space Agency (ESA) mission \emph{Gaia} (\url{https://www.cosmos.esa.int/gaia}), processed by the \emph{Gaia} Data Processing and Analysis Consortium (DPAC, \url{https://www.cosmos.esa.int/web/gaia/dpac/consortium}). Funding for the DPAC has been provided by national institutions, in particular the institutions participating in the \emph{Gaia} Multilateral Agreement.

\bibliographystyle{mnras}
\bibliography{sim}
\label{lastpage}
\end{document}